\documentclass[onecolumn,draftcls,12pt]{IEEEtran}

\usepackage{amsfonts}
\usepackage{mathrsfs}
\usepackage{amssymb,amsmath}
\usepackage[noblocks]{authblk}
\usepackage{multirow,booktabs}
\usepackage{algorithm}
\usepackage{algorithmic}
\usepackage{epsfig,graphicx}
\usepackage{stfloats}
\usepackage{theorem}
\usepackage[dvipsnames]{xcolor}
\usepackage[compress]{cite}
\usepackage[OT1]{fontenc}
\usepackage{bm}

\allowdisplaybreaks[4]

\usepackage{hyperref}
\hypersetup{
	colorlinks=true,
	linkcolor=blue,
	citecolor=blue,
}

\usepackage{subfigure}

\usepackage{pifont}
\usepackage[normalem]{ulem}

\newcommand{\T}{\mathrm{T}}
\newcommand{\hem}{\mathrm{H}}
\newcommand{\tr}{\mathrm{Tr}}

\newcommand{\diag}{\mathrm{diag}}
\newcommand{\Diag}{\mathbf{diag}}
\newcommand{\vecm}{\mathrm{vec}}

\begin{document}

\title{Dual-Functional MIMO Beamforming Optimization for RIS-Aided Integrated Sensing and Communication}

\author{
	\large
	Xin~Zhao,  Heng~Liu, Shiqi~Gong, Xin Ju, Chengwen~Xing, and  Nan~Zhao
	
	
	\thanks{
		\vspace{-5mm}
		
		Xin Zhao,  Xin Ju and Chengwen Xing  are with the School of Information and Electronics, Beijing Institute of Technology, Beijing 100081, China (e-mail: xinzhao.eecs@gmail.com;  xinjubit@gmail.com; chengwenxing@ieee.org).
		
		Heng Liu and Shiqi Gong are with the School of Cyberspace Science and Technology, Beijing Institute of Technology, Beijing 100081, China (e-mail: heng\_liu\_bit\_ee@163.com; gsqyx@163.com).
		
		Nan Zhao is with the School of Information and Communication Engineering, Dalian University of Technology, Dalian 116024, China (e-mail: zhaonan@dlut.edu.cn).
	}
	
}

\maketitle

\vspace{-16mm}

\begin{abstract}
	Aiming at providing wireless communication systems with environment-perceptive capacity, emerging integrated sensing and communication (ISAC) technologies face multiple difficulties, especially in balancing the performance trade-off between the communication and radar functions.
	In this paper, we introduce a reconfigurable intelligent surface (RIS) to assist both data transmission and target detection in a dual-functional ISAC system. To formulate a general optimization framework, diverse communication performance metrics have been taken into account including famous capacity maximization and mean-squared error (MSE) minimization. Whereas the target detection process is modeled as a general likelihood ratio test (GLRT) due to the practical limitations, and the monotonicity of the corresponding detection probability is proved. For the single-user and single-target (SUST) scenario, the minimum transmit power of the ISAC transceiver has been revealed. By exploiting the optimal conditions of the BS design, we validate that the BS is able to realize the maximum power allocation scheme and derive the optimal BS precoder in a semi-closed form. Moreover, an alternating direction method of multipliers (ADMM) based RIS design is proposed to address the optimization of unit-modulus RIS phase shifts. For the sake of further enhancing computational efficiency, we also develop a low-complexity RIS design based on Riemannian gradient descent.
	Furthermore, the ISAC transceiver design for the multiple-users and multiple-targets (MUMT) scenario is also investigated, where a zero-forcing (ZF) radar receiver is adopted to cancel the interferences in the echo signals. Then optimal BS precoder is derived under the maximum power allocation scheme, and the RIS phase shifts can be optimized by extending the proposed ADMM-based RIS design algorithm.
	Numerical simulation results verify the convergence and superior communication/sensing performance of our proposed transceiver designs.
\end{abstract}


\section{Introduction}
\label{sec:intro}

\vspace{-3mm}

Over the past decades, wireless communication and radio detection have played central roles in the development of radio technologies, which led to the flourishing of the information revolution and also greatly shaped our modern society \cite{GaoZhen2022ISAC_CS,Saad_6G_Vision}. For quite a long time, these two fundamental technologies have evolved separately due to the limitation of disjoint radio frequencies, different hardware architectures, and distinct signal waveforms. However, with the advent of the artificial intelligence (AI) era, AI-empowered devices, e.g. autonomous vehicles, not only require high-speed data transmission capacity, but also demand the capability of efficiently perceiving and sensing the surrounding environment \cite{Saad_6G_Vision,AutonomousRadar,Heath2016Sensing}. 
Motivated by the rising sensing and communication demands, integrated sensing and communication (ISAC) gathered significant attention recently, which would be able to open up new possibilities for economical manufacturing cost, enhanced system performance, and higher spectrum utilization. As a result, the ISAC topic has quickly become one of the most prominent research focuses in both the wireless communication and radar detection field.

The concept of radar and communication systems integration has a long history, dating back to the 1960s, where radar pulses were first utilized to assist the one-way data transmission task \cite{Mealey1963ISAC}. 
The subsequent study investigated the integration problem from a system level and devised a full-featured ISAC transceiver operating either in a radar mode for space rendezvous or a relay mode for high quality two-way communication \cite{CagerTCOM1978Orbiter}. 
Although the ISAC design can bring the integration gain and the coordination gain from the reuse of resources, the corresponding investigations have been trapped in stagnation for years \cite{LiuFan2022Survey}. One of the most important reasons is that wireless technologies were not powerful enough to remedy the difficulties of joint communication and radar designs until the emergence of multiple-input and multiple-output (MIMO) technologies. Driven by the demand for high-speed transmission and high precision detection, respectively, both the communication and radar systems have been evolving toward the direction of high-frequency bands and large-scale antenna arrays\cite{Xing2020Part1,Xin2022TWC,MIMORadar2006TSP,PhaseMIMORadar2010TSP}. This convergence trend bridges the gaps between the two systems. Further, the MIMO arrays also grant transceivers extra degrees-of-freedoms (DoFs) to flexibly optimize the trade-off between the radar and communication performance 
\cite{Stoica2007MIMORadar,SpatialDiversityMIMORadar,XingTSP2019Hybrid}. Therefore, a plethora of studies have emerged focusing on the joint radar and communication systems designs in various ISAC scenarios, including enhanced positioning, in-door object sensing, simultaneous localization and mapping (SLAM), etc \cite{LiuFan2022Survey}.

To implement radar and communication integration, the radar and communication in an ISAC system often share certain hardware resources, such as antennas, radio-frequency (RF) processors, AD/DA converters, etc\cite{ZhangAndrew2022Survey}. Based on this idea, dual-function radar-communication (DFRC) systems are proposed where the radar and communication functions are combined in one system architecture to reduce the redundancy of hardware and improve spectral efficiency \cite{LiuFan2018DFRC,Eldar2020DFRC}.
Early studies focused on the joint ISAC system design with a single-antenna DFRC transceiver. In \cite{SingleAntennaDFRC2016}, the time-division multiple access technique was adopted to isolate the signals for different functions. These schemes suffer from severe performance loss due to the limited DoFs in the antenna domain \cite{Eldar2020TSP}. Consequently, in the following research, the waveform diversity has been taken into consideration and data symbols were embedded into the radar waveform to achieve concurrent data transmission during radar sensing \cite{SidelobeTSP2015Dual}. However, the data transmission rate is limited by the radar pulse frequency in such information embedding systems. Further, the transmit beamforming based ISAC system architecture was proposed in a point-to-point communication scenario, where the radar and communication signal processing tasks are explicitly incorporated in the beamformer \cite{LiuFan2018DFRC}. Then, the ISAC system design was investigated for the multi-user communication scenario in \cite{Eldar2020TSP}, where individual radar and communication waveforms were optimized and transmitted from the same DFRC transmitter.
Note that the shared resources in a DFRC transceiver inevitably lead to the overlapping of sensing and communication signals in either time or frequency domain \cite{Co-Design2016TSP}. These individual waveform designs generally required beam pattern management to suppress the interference \cite{Eldar2020TSP,LiuFan2018MU-MIMO}. Inspired by the concept of passive radar, a novel joint communication and radar beamforming design were proposed, where the radar target was illuminated by the carefully devised communication signals \cite{LiuFan2021JointDFRC,ZhangAndrew2021JSTSP}. Therefore, the ISAC systems are able to perform better trade-offs between the radar and communication counterparts \cite{GeneralizedDFRC2022JSAC}.

On the other hand, as one of the key enabling technologies for the next generation of wireless systems, the reconfigurable intelligent surface (RIS) is able to provide not only massive connections and ultra-reliable transmission for communications, but also the enhanced target detection and parameter sensing accuracy \cite{GongTSP2021Unified,Xin2022TCOM,RIS-MIMORadar2022TSP}. In \cite{RIS-MIMORadar2022TSP}, the authors investigated the target detection optimization in a MIMO radar system with multiple RISs. Motivated by the success in both communication and radar fields, the RIS technique was also introduced in the ISAC system, which could provide extra spatial DoFs to further enhance either the transmission or the detection performance \cite{ISAC-RIS2023Magazine}. The initial studies only regarded the RIS as an additional component for the communication purpose and the radar sensing was not affected by the RIS \cite{WangXY2021TVT,RIS-ISAC2022TVT,HuaMeng2023ISAC}. However, these designs did not fully capitalize on the diversity offered by the RIS 
\cite{YunlongCai2022RIS-ISAC,YuanXJ2022Passive,ZhangRui2022IRS-ISAC,ZhangHL2022COML}. To assist the sensing process, a self-sensing RIS architecture was proposed in \cite{ZhangRui2022IRS-ISAC}, where the RIS consists of phase shifters and sensors. Recently, there are also some research works focusing on the ISAC system design with a passive RIS, and a semi-definite relaxtion (SDR) based design was presented to overcome the non-convex radar detection constraint \cite{YuanXJ2022Passive}. The double-RIS-aided ISAC system was investigated in \cite{YunlongCai2022RIS-ISAC}, where the transceiver was optimized by maximizing the communication signal-to-interference-plus-noise ratio (SINR). To unify the radar and communication beamforming designs, the ISAC transceiver design was formulated as a weighted mutual information maximization problem in \cite{ZhangHL2022COML}.

To the best of the authors' knowledge, the RIS-aided ISAC system design is still in its infancy. Most existing RIS-aided joint sensing and communication beamforming designs only considered the MISO communication scenario with a single target \cite{YuanXJ2022Passive,ZhangRui2022IRS-ISAC,ZhangHL2022COML}, the corresponding studies on MIMO transmission with multiple targets are barely discussed. Moreover, the numerous performance metrics for communications and sensing require case-by-case studies for various ISAC scenarios, which is inconvenient for practical implementation. This motivates us to carry out a unified ISAC transceiver optimization framework.

Regarding these concerns, in this paper, we investigate the general dual-functional  beamforming design in a RIS-aided ISAC system. Different from most existing ISAC related works, the  introduction of  RIS   not only facilitates the base station (BS) in broadcasting information, but also empowers it with the ability to detect potential targets in the severely blocked area. Moreover, the MIMO technology has been taken into account, being suitable for boosting both communication and sensing performance. The main contributions of our paper are summarized as follows.
\begin{itemize}
	\item We propose a general optimization framework for the RIS-aided dual-functional ISAC system design in the single-user and single-target (SUST) scenario. The framework includes DFRC transceiver designs with multiple communication performances, including capacity maximization, mean-squared error (MSE) minimization, etc. Specifically, considering the practical sensing limitations, the target sensing procedure is modeled as a general likelihood ratio test (GLRT). Moreover, the monotonicity of the probability of detection with respect to (w.r.t.) the received SINR is proved. 
	\item By exploiting the optimal condition, it is found that the BS satisfies the maximum power allocation criterion. Based on this, the BS precoder design is transformed into a quadratic constrained optimization and the optimal BS precoder can be derived in semi-closed form. Then, an iteratively ADMM-based RIS design is proposed to address the unit-modulus phase shift optimization with the quartic detection probability constraint. Furthermore, in order to circumvent the high complexity of the ADMM algorithm, a Riemannian gradient descent based RIS design is proposed for enhancing computational efficiency.
	\item Finally, based on derived general system model, the ISAC transceiver optimization problem for the multiple-users and multiple-targets (MUMT) scenario is under investigation, where a zero-forcing (ZF) radar receiver has been adopted to eliminate the negative impact of the interferences in the echo signals. Moreover, it is found that the optimal BS precoder satisfies the maximum power allocation scheme as well, and based on which, the BS precoder optimization is simplified. Further, it is shown that the general ISAC system design with multiple communication objectives can also be applied to the RIS phase shift design for the MUMT scenario.
\end{itemize}

The rest of this paper is organized as follows. We introduce the signal models for the communication and radar components in the ISAC system as well as the performance metrics, followed by the problem formulation in Section~\ref{sec:sys_model}. The transceiver optimization and computational efficient RIS design for the SUST scenarios are presented in Section~\ref{sec:single_ISAC}. Then, the ISAC system design for the MUMT scenarios is given in Section~\ref{sec:multi_ISAC}. The numerical simulations for the proposed design are shown in Section~\ref{sec:simulation} and the conclusion is drawn in Section~\ref{sec:conclusion}.

\textit{Notations:} In this paper, the following symbol notations and conventions are used without other specification. Scalars, vectors, and matrices are written as non-bold, bold lower-case, and bold upper-case letters, respectively. The symbol $ \mathbb{C} $ denotes the set of complex numbers. For the matrices operations, $ {\bf X}^{-1} $, $ {\bf X}^* $, $ {\bf X}^\hem $, $ \tr( {\bf X} ) $, and $ \vert {\bf X} \vert $ denote the inversion, conjugate, hermitian, trace, and determinant of a complex matrix $ {\bf X} $, respectively. The diagonal matrix with diagonal elements of a vector $ {\bf x} $ is given by $ \Diag( {\bf x} ) $ and the vector formed by the diagonal elements of a matrix $ {\bf X} $ is expressed as $ \diag( {\bf X}) $. The $ i $th column and $ j $th row of a matrix $ {\bf X} $ are denoted by $ [{\bf X}]_{:,i} $ and $ [{\bf X}]_{j,:} $, respectively. The matrices with its $ i $th column or $ j $th row removed are written as $ [{\bf X}]_{:,-i} $ and $ [{\bf X}]_{-j,:} $, respectively. Furthermore, the real and imaginary part of a complex number are denoted by $ \Re\{ \cdot \} $ and $ \Im\{ \cdot \} $, respectively.

\vspace{-3mm}

\section{System Model}
\label{sec:sys_model}

\vspace{-1.5mm}

We consider a general ISAC system, where a DFRC-BS serves $ K $ communication users and detects $ L $ blocked targets as depicted in Fig~\ref{fig_sys}. The BS is equipped with $ N_{\rm B} $ transmit and $ N_{\rm B} $ receive antennas, and each user is installed with $ N_{\rm U} $ antennas. 
The BS operates in a full-duplex mode to transmit data signal and receive echo signals simultaneously and the self-interference is assumed to be perfectly canceled. 
Due to the lack of line-of-sight (LoS) link between the BS and the targets, a uniform planar RIS antenna array with $ N_{\rm R} $ passive tunable elements is deployed to assist both the target detection and data transmission. In this paper, we consider the uniform linear array (ULA) at the BS/each user and the uniform planar array (UPA) at the RIS.


\vspace{-6mm}

\subsection{Signal Models}

\begin{figure}[!t]
	\vspace{-7mm}
	\centering
	\subfigure[The system model diagram for the RIS-aided ISAC system.]{
		\centering
		\includegraphics[width=0.50\textwidth, trim={5mm 5mm 5mm 5mm}]{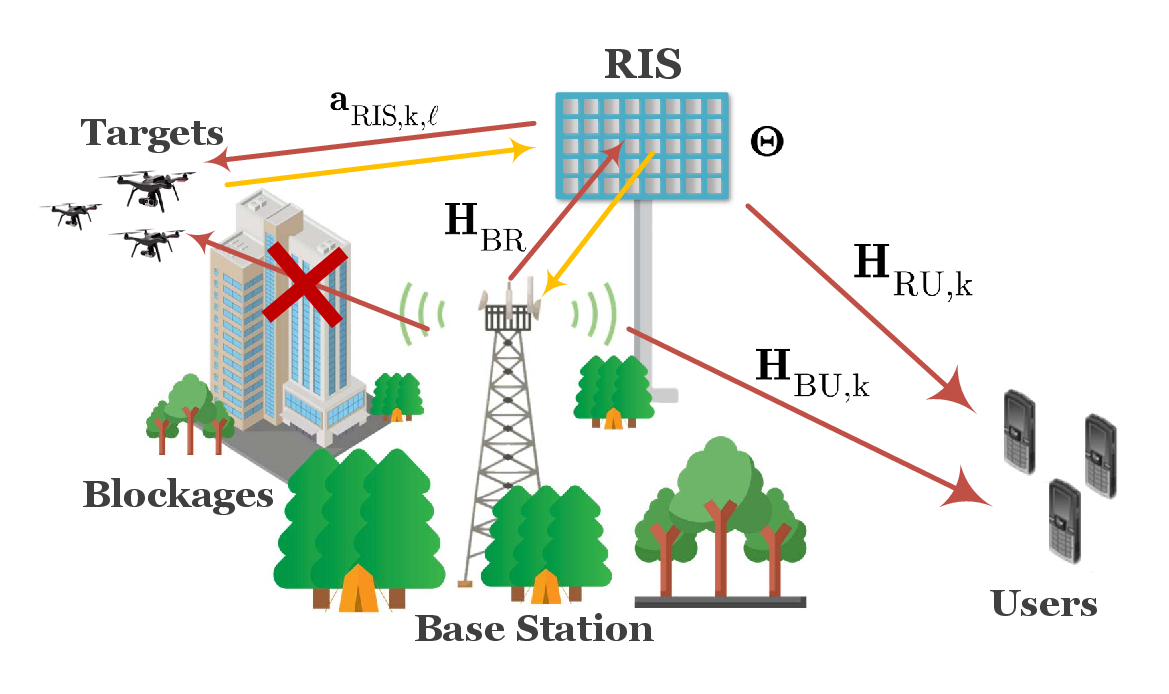}
		\vspace{-4mm}
		\label{fig_sys}
	}
	\hfill
	\subfigure[Illustration for time slot allocation of the ISAC system.]{
		\centering
		\includegraphics[width=0.43\textwidth, trim={4mm 0mm 7mm 0mm}]{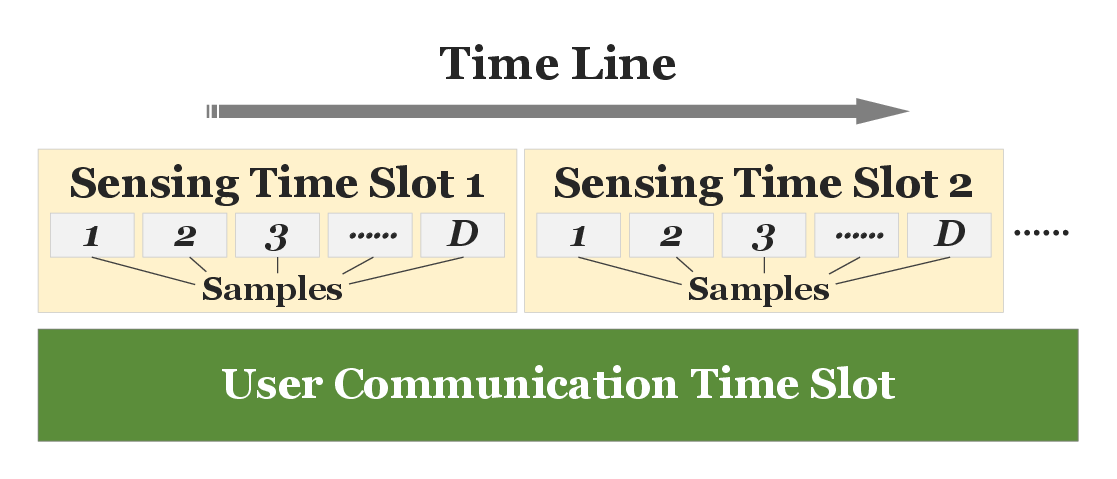}
		\vspace{-5mm}
		\label{fig_time_slot}
	}
	\vspace{-2mm}
	\caption{The system model set-up and joint communication and sensing scheme for the ISAC system.}
	\label{fig_sys_model_all}
	\vspace{-10mm}
\end{figure}

\vspace{-2mm}

For the communication subsystem, the BS precodes the data symbols $ {\bf s}_k \in \mathbb{C}^{N \times 1} $ for user $ k $ with a beamformer $ {\bf F}_k \in \mathbb{C}^{N_{\rm B} \times N} $ and broadcasts the precoded signal to users. Then the received signal $ {\bf y}_{{\rm c},k} \in \mathbb{C}^{N_{\rm U} \times 1} $ at user $ k $ can be modeled as
\begin{align}
	{\bf y}_{{\rm c},k} & = ( {\bf H}_{{\rm BU},k} + {\bf H}_{{\rm RU},k} \boldsymbol{\Theta} {\bf H}_{\rm BR}^\hem ) {\bf F} {\bf s} + {\bf n}_{{\rm c},k} 
	\label{Eq_y_MU}
\end{align}
where $ \boldsymbol{\Theta} = \Diag \{ [\theta_1, \ldots, \theta_{N_{\rm R}}] \} $ is the phase shift matrix of the RIS. $ {\bf H}_{{\rm BU},k} \in \mathbb{C}^{N_{\rm U} \times N_{\rm B}} $, $ {\bf H}_{{\rm RU},k} \in \mathbb{C}^{N_{\rm U} \times N_{\rm R}} $, $ {\bf H}_{\rm BR}^\hem \in \mathbb{C}^{N_{\rm R} \times N_{\rm B}} $ represent the channel matrices between the BS and user $ k $, the RIS and user $ k $, and the BS and RIS, respectively. In addition, $ {\bf F} = [ {\bf F}_1, \ldots, {\bf F}_K ] $ and $ {\bf s} = [ {\bf s}_1^\T, \ldots, {\bf s}_K^\T ]^\T $ are the stacked precoders and data streams, respectively. The vector $ {\bf n}_{{\rm c},k} \in \mathbb{C}^{N_{\rm R} \times 1} $ denotes the complex white Gaussian noise with zero mean and covariance $ {\bf R}_{{\bf n}_{\rm c}} $.
Then, the MSE matrix of the received signal at user $ k $ is formulated as
\begin{align}
	\boldsymbol{\Phi}_{{\rm MSE},k}
	& \! = \! ( {\bf G}_k {\bf H}_{{\rm c},k} {\bf F}_k \! - \! {\bf I} ) ( {\bf G}_k {\bf H}_{{\rm c},k} {\bf F}_k \! - \! {\bf I} )^\hem \! + \! {\bf G}_k \widetilde{\bf R}_{{\bf n}_{{\rm c},k}} {\bf G}_k^\hem.
	\label{Eq_MSE_MU}
\end{align}
where $ \widetilde{\bf R}_{{\bf n}_{{\rm c},k}} \triangleq \sum\nolimits_{\ell \neq k} {\bf H}_{{\rm c},k} {\bf F}_{\ell} {\bf F}_{\ell}^\hem {\bf H}_{{\rm c},k}^\hem \! + \! {\bf R}_{{\bf n}_{\rm c}} $, and the effective channel is defined as $ {\bf H}_{{\rm c},k} \triangleq {\bf H}_{{\rm BU},k} + {\bf H}_{{\rm RU},k} \boldsymbol{\Theta} {\bf H}_{\rm BR}^\hem $ and $ {\bf G}_k $ is the receive equalizer for user $ k $.

In contrast, the radar subsystem utilizes transmitted communication signals, i.e., $ {\bf x} = {\bf F} {\bf s} $, to sense remote targets. As the locations of the targets are unknown in prior, a sweeping sensing policy is adopted as in \cite{YuanXJ2022Passive}. The BS generates a sensing beam towards the chosen direction in each time slot and moves the beam to another direction in the next time slot. Thus, the received echo signal at the radar receiver is formulated as \cite{YuanXJ2022Passive}
\begin{align} 
	{\bf y}_{\rm r} \!\! = \!\! \sum_{\ell = 1}^L \!\! \frac{ \lambda^2 }{ ( 4 \pi d_{{\rm RT},\ell} )^2 } 
	{\bf H}_{\rm BR} \boldsymbol{\Theta} {\bf a}_{\rm RIS,r}( \boldsymbol{\beta}_{\ell} ) \!
	\bigg( \!\! \iint\limits_{S_{\ell}} \!\! \frac{4 \pi}{\lambda^2} \! \big| g_{\rm T}( x, y ) \big|^2 \! dx dy \! \bigg)^{\!\!\!1/2} \!\!\!
	{\bf a}_{\rm RIS,t}^\hem( \boldsymbol{\beta}_{\ell} ) \boldsymbol{\Theta} {\bf H}_{\rm BR}^\hem {\bf x}
	\! + \! {\bf n}_{\rm r},
\end{align}
where $ \lambda $ is the wavelength of the signal, $ d_{{\rm RT},\ell} $ is the distance between the $ \ell $th target and the RIS, and $ {\bf n}_{\rm r} $ stands for the complex white Gaussian noise vector with zero mean and covariance $ {\bf R}_{{\bf n}_{\rm r}} $.
$ g_{\rm T}( x, y ) $ is referred to as the reflective radiation power density at the coordinate $ (x, y) $ and $ S_\ell $ is the scattering surface area.
The vectors $ {\bf a}_{\rm RIS,t}( \boldsymbol{\beta}_{\ell} ) $ and $ {\bf a}_{\rm RIS,r}( \boldsymbol{\beta}_{\ell} ) $ denote the transmit and receive steering vectors of the RIS, where $ \boldsymbol{\beta} \!=\! \{ \vartheta_{\ell}, \phi_{\ell} \} $ is azimuth and elevation angles set for the RIS-to-$ \ell $th-target link, respectively. For notational simplicity, the corresponding steering vectors are simplified to $ {\bf a}_{{\rm RIS,t},\ell} $ and $ {\bf a}_{{\rm RIS,r},\ell} $ in the following parts.

Define matrices $ {\bf A}_{\rm RIS,r} \! = \! [ {\bf a}_{{\rm RIS,r},1}, \ldots, {\bf a}_{{\rm RIS,r},L} ] $, $ {\bf A}_{\rm RIS,t} \! = \! [ {\bf a}_{{\rm RIS,t},1}, \ldots, {\bf a}_{{\rm RIS,t},L} ] $, and the coefficient matrix $ \boldsymbol{\Xi} \! = \! \diag\big( [ \rho_{{\rm tg},1}, \ldots, \rho_{{\rm tg},L} ] \big) $, where $ \rho_{{\rm tg},\ell} \! = \! \frac{ \lambda }{ ( 4 \pi )^{3/2} d_{{\rm RT},\ell}^2 } \sqrt{ \! \iint\nolimits_{S_{\ell}} \! \big| g_{{\rm T},\ell}( x, y ) \big|^2 dx dy } $ is the reflection coefficient of the $ \ell $th target.
The detected signal at the radar receiver is expressed as
\begin{align}
	\hat{\bf y}_{\rm r} = {\bf W}^\hem {\bf y}_{\rm r}
	= {\bf W}^\hem {\bf H}_{\rm BR} \boldsymbol{\Theta} {\bf A}_{\rm RIS,r} \boldsymbol{\Xi} {\bf A}_{\rm RIS,t}^\hem \boldsymbol{\Theta} {\bf H}_{\rm BR}^\hem {\bf x}
	+ {\bf W}^\hem {\bf n}_{\rm r},
	\label{Eq_Detected_Sig_Matrix}
\end{align}
where $ {\bf W} = [ {\bf w}_1, \ldots, {\bf w}_L ] $ denotes the stacked radar receive beamformers. As is shown in Fig~\ref{fig_time_slot}, to improve the sensing accuracy, the radar receiver considers $ D $ samples in each sensing slot to determine the existence of a target. Then, the sampled signal for the $ \ell $th target is written as
\begin{align}
	\hat{y}_{{\rm r},\ell}[n] & = \rho_{{\rm tg},\ell} {\bf w}_{\ell}^\hem {\bf H}_{\rm BR} \boldsymbol{\Theta} {\bf a}_{{\rm RIS,r},\ell} {\bf a}_{{\rm RIS,t},\ell}^\hem \boldsymbol{\Theta} {\bf H}_{\rm BR}^\hem {\bf x}[n] 
	+ {\bf w}_{\ell}^\hem {\bf n}_{\rm r}[n] \notag \\
	& + \sum_{j \neq \ell} \rho_{{\rm tg},j} {\bf w}_{\ell}^\hem {\bf H}_{\rm BR} \boldsymbol{\Theta} {\bf a}_{{\rm RIS,r},j} {\bf a}_{{\rm RIS,t},j}^\hem \boldsymbol{\Theta} {\bf H}_{\rm BR}^\hem {\bf x}[n],
	 \ \, \forall n = 1, \! \ldots, \! D,
	\label{Eq_Detected_Sig_MT}
\end{align}
where the first term denotes the desired echo from the $ k $th target, and the third term is the interference signals from other targets. 
Based on the received signal (\ref{Eq_Detected_Sig_MT}), the radar is able to determine the existence of the targets.

\vspace{-5mm}

\subsection{Communication Performance}
\label{sec:sys_model:comm_perform}

\vspace{-2mm}

For communication systems, different choices of performance metrics generally lead to distinct final behaviors. Fortunately, it is revealed that most of them can be expressed as Schur-concave functions w.r.t. the MSE matrix, i.e., \cite{Xing2020Part1,Xin2022TCOM}
\begin{align}
	f_{\rm obj}( \cdot ) \triangleq \sum_{k = 1}^K f_{\rm Concave}^{\rm{Schur}} \big[ {\rm diag} \big( \boldsymbol{\Phi}_{{\rm MSE},k} \big) \big].
\end{align}
The objective functions generally satisfy $ \frac{\partial f_{\rm obj} ( {\bf x} )}{\partial x_i} > 0 $, which indicates that the system performance is enhanced with more transmit power. Among numerous performance metrics, the weighted sum rate and the weighted MSE performance metrics are the most representative. Thus, we take them as two typical examples to illustrate the Schur-concave objectives.

\textit{1) Weighted MSE Performance}
is a popular metric for reliable transmissions, i.e.,
\begin{align}
	f_{\rm wmse}( \cdot ) \triangleq \sum_{k = 1}^K \tr \big( {\bf A}_k^\hem \boldsymbol{\Phi}_{{\rm MSE},k} {\bf A}_k \big),
	\label{Obj_WMSE}
\end{align}
where $ {\bf A}_k $ is a weighting matrix for user $ k $. Specifically, considering the single-user case with an identity weighting matrix, the objective function (\ref{Obj_WMSE}) downgrades to the ordinary MSE performance, i.e.,
$ f_{\rm mse}( \cdot ) \triangleq \tr \big\lbrace ( {\bf G} {\bf H}_{\rm c} {\bf F} \! - \! {\bf I} ) ( {\bf G} {\bf H}_{\rm c} {\bf F} \! - \! {\bf I} )^\hem \! + \! {\bf G} {\bf R}_{{\bf n}_{\rm c}} {\bf G}^\hem \big\rbrace $.

\textit{2) Weighted Sum Rate Performance}
is the most important measurement for the throughput capability of a communication system, which can be expressed in a general form, i.e., \cite{Xing2020Part1}
\begin{align}
	f_{\rm wsr}( \cdot ) \triangleq - \sum_{k = 1}^K \log \big| \boldsymbol{\Phi}_k + {\bf C}_k^\hem {\bf F}_k^\hem {\bf H}_{{\rm c},k}^\hem \widetilde{\bf R}_{{\bf n}_{{\rm c},k}}^{-1} {\bf H}_{{\rm c},k} {\bf F}_k {\bf C}_k \big|,
	\label{Obj_WMI}
\end{align}
where $ \boldsymbol{\Phi}_k $ and $ {\bf C}_k $ are a positive semi-definite matrix and a weighting matrix for user $ k $, respectively.
Assume both $ \boldsymbol{\Phi}_k $ and $ {\bf C}_k $ to be identity matrices, then the objective function (\ref{Obj_WMI}) downgrades to the famous capacity metric, i.e.,
$ f_{\rm sr}( \cdot ) \triangleq - \log \big| {\bf I} + {\bf F}^\hem {\bf H}_{\rm c}^\hem {\bf R}_{{\bf n}_{\rm c}}^{-1} {\bf H}_{\rm c} {\bf F} \big| $.

\vspace{-5mm}

\subsection{Problem Formulation}
\label{sec:sys_model:problem}

\vspace{-2mm}

According to the integration property of the ISAC system, the transceiver design is naturally separated into the radar and communication subproblems. It is seen from the radar received signal~(\ref{Eq_Detected_Sig_MT}) that the interference has a non-negligible influence on the correct detection of a target. A viable criterion for the radar receive beamforming is to make the power of the desired echo signal as large as possible and that of the interference as small as possible simultaneously. Thus, the radar receive beamforming design can be restated as
\begin{align}
	\max_{{\bf w}_{\ell}} \ & \frac{ \Vert {\bf F}^\hem {\bf H}_{\rm BR} \boldsymbol{\Theta}^\hem {\bf a}_{{\rm RIS,t},\ell} {\bf a}_{{\rm RIS,r},\ell}^\hem \boldsymbol{\Theta}^\hem {\bf H}_{\rm BR}^\hem {\bf w}_{\ell} \Vert_2^2 }
	{ {\bf w}_{\ell}^\hem {\bf w}_{\ell} } \notag \\
	{\rm s.t.} \ & \Big\Vert \sum\nolimits_{j \neq \ell} \rho_{{\rm tg},\ell} {\bf w}_{\ell}^\hem {\bf H}_{\rm BR} \boldsymbol{\Theta} {\bf a}_{{\rm RIS,r},j} {\bf a}_{{\rm RIS,t},j}^\hem \boldsymbol{\Theta} {\bf H}_{\rm BR}^\hem {\bf F} \Big\Vert_2^2 \le \epsilon,
	\label{Radar_Beamformer_Opt_Prob}
\end{align}
where $ \epsilon $ is a given small threshold. To better explore the performance trade-off between the radar and communication subsystems, we limit our study to the case that the radar beamformer perfectly cancels the interference, i.e., $ \epsilon = 0 $. 
Based on the detected signal, the target sensing process can be modeled as binary hypothesis tests \cite{DetectionTheory,YuanXJ2022Passive}. Generally, the echoes of targets are deterministic but not perfectly known at the radar. In this case, the GLRT is utilized, where the unknown parameters are substituted by their maximum likelihood estimators (MLEs). According to the detection theory \cite{DetectionTheory}, the probability of detection for the $ \ell $th target can be modeled as
\begin{align}
	P_{D,\ell}
	= Q \Bigg( Q^{-1} \bigg( \frac{ P_{FA} }{2} \bigg) - \sqrt{ \frac{ \mathcal{E} }{ \Vert {\bf w}_{\ell} \Vert_2^2 \sigma_{n,r}^2 } } \, \Bigg)
	+ Q \Bigg( Q^{-1} \bigg( \frac{ P_{FA} }{2} \bigg) + \sqrt{ \frac{ \mathcal{E} }{ \Vert {\bf w}_{\ell} \Vert_2^2 \sigma_{n,r}^2 } } \, \Bigg),
	\label{Radar_Constraint_Def}
\end{align}
where $ P_{FA} $ is the probability of false alarm and $ \mathcal{E} $ is the energy of the detected signal.
With the communication and radar performance metrics at hand, we formulate the ISAC system optimization problem as follows according to the radar centric criterion, i.e.,
\begin{subequations}
	\label{General_Opt_Model}
	\begin{align}
		\min_{\substack{ \{ {\bf F}_k \}, \boldsymbol{\Theta}, \{ {\bf G}_k \} } } \ 
		& \sum\limits_{k = 1}^K f_{\rm obj} \big( \boldsymbol{\Phi}_{{\rm MSE},k} \big) \\
		{\rm s.t.} \ \ \ \ \ & \sum\limits_{k = 1}^K \tr( {\bf F}_k {\bf F}_k^\hem ) \le P_{\rm max}, 
		\label{General_Opt_Model:C1} \\
		& P_{D,\ell} \ge \gamma_{\rm D}, \ \ \forall \ell = 1, \ldots, L, 
		\label{General_Opt_Model:C2} \\
		& | \theta_i | = 1, \ \ i = 1, \ldots, N_{\rm R},
		\label{General_Opt_Model:C3}
	\end{align}
\end{subequations}
where (\ref{General_Opt_Model:C1}) denotes the total power constraint of the BS, (\ref{General_Opt_Model:C2}) guarantees the detection performance, and (\ref{General_Opt_Model:C3}) models the unit-modulus property of RIS elements. It is worth noting that the optimization problem (\ref{General_Opt_Model}) is formulated from a communication-centric perspective, which means that the BS incorporates the sensing function to a communication system, and the data transmission performance is concerned with top priority.

The problem (\ref{General_Opt_Model}) describes a general optimization model for a multi-user transmission and multi-target sensing ISAC system. However, the complexity of this model generally hinders the revelation of the intrinsic nature of the ISAC system. In the following sections, we firstly investigate the ISAC system design under the SUST scenario, and then extend the proposed algorithm to the complicated MUMT scenario.

\vspace{-4mm}

\section{ISAC System Optimization For the SUST Scenarios}
\label{sec:single_ISAC}

\vspace{-2mm}

In this section, we consider the interference-free scenario, where there is only a single-user in the cell and one potential sensing target. By annulling the interference signals, we have an opportunity to clarify the trade-off relationship between the radar and communication subsystems and propose a joint transceiver design to achieve the optimal performance. 

\vspace{-6mm}

\subsection{Optimal Radar Receive Beamforming Design}

\vspace{-2mm}

We firstly investigate the radar receive beamforming optimization problem with the BS precoder and the RIS fixed. Since there is no interference from other targets, the radar beamforming design (\ref{Radar_Beamformer_Opt_Prob}) is simplified to
\begin{align}
	\max_{\bf w} \ & \frac{ \Vert {\bf F}^\hem {\bf H}_{\rm BR} \boldsymbol{\Theta}^\hem {\bf a}_{\rm RIS,t} {\bf a}_{\rm RIS,r}^\hem \boldsymbol{\Theta}^\hem {\bf H}_{\rm BR}^\hem {\bf w} \Vert_2^2 }
	{ {\bf w}^\hem {\bf w} },
	\label{Radar_Opt_1}
\end{align}
which is a typical Rayleigh quotient optimization, and the optimal radar receive beamformer is given by 
\begin{align}
	\label{Opt_w_ST}
	{\bf w}_{\rm opt} = \beta_{\rm r} {\bf H}_{\rm BR} \boldsymbol{\Theta} {\bf a}_{\rm RIS,r},
\end{align}
where $ \beta_{\rm r} = \frac{1}{\Vert {\bf H}_{\rm BR} \boldsymbol{\Theta} {\bf a}_{\rm RIS,r} \Vert_2} $ is a scalar to control the power of the detected signal.

\vspace{-5mm}

\subsection{Optimal DFRC-BS Precoder Design}
\label{sec:single_ISAC:BS}

\vspace{-2mm}

After obtaining the optimal radar receiver, we embark on the transceiver design for the DFRC-BS.
Recalling problem (\ref{General_Opt_Model}), the BS precoder optimization problem can be expressed as 
\begin{subequations}
	\label{Opt_1}
	\begin{align}
		\min_{ {\bf F}, {\bf G} } \ & f_{\rm obj} \big( \boldsymbol{\Phi}_{\rm MSE} \big) \\
		{\rm s.t.} \ \ & \tr( {\bf F} {\bf F}^\hem ) \le P_{\rm max}, 
		\label{Opt_1:C} \\
		& P_{D} \ge \gamma_{\rm D}.
		\label{Opt_1:C2}
	\end{align}
\end{subequations}
The detection probability constraint differentiates the ISAC system design (\ref{Opt_1}) from the traditional one in \cite{Xin2022TCOM}, and the two main challenges come from the non-convexity of the radar constraint and the entangled radar and power constraints. In order to reveal the optimal structure of the BS, it is urgent to investigate the radar constraint firstly.
According to the signal model in (\ref{Radar_Constraint_Def}), the energy of the detected signal is written as $ \mathcal{E} = \tr ( {\bf F}^\hem {\bf B} {\bf F} ) $,
where $ {\bf B} $ is the covariance matrix of the radar steering vector
\begin{align}
	{\bf B} = \, & \big| \rho_{\rm tg} \big|^2 {\bf H}_{\rm BR} \boldsymbol{\Theta}^\hem {\bf a}_{\rm RIS,t} {\bf a}_{\rm RIS,r}^\hem \boldsymbol{\Theta}^\hem {\bf H}_{\rm BR}^\hem {\bf w} 
	{\bf w}^\hem {\bf H}_{\rm BR} \boldsymbol{\Theta} {\bf a}_{\rm RIS,r} {\bf a}_{\rm RIS,t}^\hem \boldsymbol{\Theta} {\bf H}_{\rm BR}^\hem.
	\label{Def_B}
\end{align}
Notice that the probability function in (\ref{Radar_Constraint_Def}) is non-analytic and non-convex. To deal with the non-convexity, we propose the following proposition.

\noindent\textit{\textbf{Proposition 1}}: Given a real scalar $ a > 0 $, we define the function $ P(x) = Q( \sqrt{x} + a ) + Q( a - \sqrt{x} ) $, then the inverse function of $ P(x) $, i.e., $ P^{-1} (x) $, exists and is strictly monotonically increasing w.r.t. $ x $.

\textit{Proof:} The proof is relegated to Appendix~\ref{sec:appendix-A}.

Based on Proposition~1, let $ a = Q^{-1} \big( \frac{ P_{FA} }{2} \big) $, then the radar constraint (\ref{Opt_1:C2}) is transformed into
\begin{align}
	\tr \big( {\bf F}^\hem {\bf B} {\bf F} \big) \ge \bar{\gamma}_{\rm D},
	\label{Radar_quadratic}
\end{align}
where the effective detection threshold $ \bar{\gamma}_{\rm D} $ is defined as $ \bar{\gamma}_{\rm D} \triangleq \frac{ \Vert {\bf w} \Vert_2^2 \sigma_{n,r}^2 P^{-1} ( \gamma_{\rm D} ) }{ D } $ for notational simplicity. Notice that the non-convexity of the radar constraint is successfully removed and the equivalent constraint (\ref{Radar_quadratic}) is quadratic.

Subsequently, we move our attention to the transceiver optimization problem (\ref{Opt_1}). Due to the free-of-constraint for the equalizer $ {\bf G} $, its optimal solution can be directly derived based on the least minimal mean-squared error (LMMSE) criterion \cite{Xin2022TCOM}.
For the precoder design, the optimization can be very tricky, because the power constraint and radar constraint (\ref{Radar_quadratic}) are coupled and the radar constraint is generally non-convex \cite{Boyd04}. To tackle the difficulties, we first propose the following proposition.

\noindent\textit{\textbf{Proposition 2}}: 
For the optimal precoder $ {\bf F}_{\rm opt} $ in (\ref{Opt_1}), the maximum power allocation scheme must be achieved.

\textit{Proof:} See Appendix~\ref{sec:appendix-B}.

Proposition 2 is easy to be understood since the extra power can always be used to enhance the communication or detection performance. Before optimizing the BS precoder, there is still a question about the minimal transmit power required by the ISAC system. 

\subsubsection{\textbf{The Minimum Transmit Power For Sensing}}

To answer this question, we consider the insufficient transmit power case. Notice that the detection demand must be satisfied in the first place, the minimum transmit power optimization is formulated as
\begin{align}
	\min_{ {\bf F} } \ \tr( {\bf F}{\bf F}^\hem ), \ \ 
	{\rm s.t.} \, \tr \big( {\bf F}^\hem {\bf B} {\bf F} \big) \ge \bar{\gamma}_{\rm D}.
	\label{Opt_Min_P}
\end{align}
By diagonalizing the matrix $ {\bf B} $, it is found that the minimum transmit power just validates the radar constraint. The optimal solution of (\ref{Opt_Min_P}) is derived as $ {\bf F} = {\bf U}_{\rm B} \boldsymbol{\Sigma}_{\rm F} $, where $ {\bf B} = {\bf U}_{\rm B} \boldsymbol{\Sigma}_{\rm B} {\bf U}_{\rm B}^\hem $, and the power allocation is given by $ [ \boldsymbol{\Sigma}_{\rm F} ]_{ii} = \sqrt{ \frac{ \bar{\gamma}_{\rm D} }{ [\boldsymbol{\Sigma}_{\rm B}]_{ii}} } $, if $ i = \arg \max_t \{ [\boldsymbol{\Sigma}_{\rm B}]_{tt} \} $ and $ [ \boldsymbol{\Sigma}_{\rm F} ]_{ii} = 0 $, otherwise.
Therefore, the minimum transmit power for the DFRC-BS is given by
\begin{align}
	P_{\rm min} = \frac{ \bar{\gamma}_{\rm D} }{ \max \{ [\boldsymbol{\Sigma}_{\rm B}]_{tt} \}}.
	\label{Minimal_Power}
\end{align}

\subsubsection{\textbf{Precoder Design for the DFRC-BS}}

With the knowledge of the minimum required power above, we focus on the precoder optimization under the sufficient BS transmit power assumption. Concerning the numerous performance metrics, the problem (\ref{Opt_1}) adopts the general Schur-concave function as the objective.
We notice that the MSE matrix plays a central role in the Schur-concave performance metrics. To optimize the general problem, a viable solution is to transform the general precoder design into a maneuverable MSE-related form.

The weighted MSE performance is naturally a quadratic function w.r.t. the BS precoder. However, the weighted sum rate optimization would be quite different. To overcome the difficulties, we extend the famous capacity--MSE equivalence in \cite{QingjiangShi2013TSP} to a general case and propose the following proposition.

\noindent\textit{\textbf{Proposition 3:}}
Given a positive definite matrix $ \boldsymbol{\Phi} $ and a matrix set $ \mathcal{X} $, the weighted sum rate maximization problem in the following form
\begin{align}
	\max_{ {\bf X} } \ \, 
	- \log \big| \boldsymbol{\Phi} + {\bf C}^\hem {\bf X}^\hem \boldsymbol{\Pi} {\bf X} {\bf C} \big|, 
	\ \ 
	{\rm s.t.} \ \ 
	{\bf X} \in \mathcal{X},
	\label{Opt_WMI}
\end{align}
is equivalent to the weighted MSE minimization below
\begin{align}
	\min_{ {\bf X}, \bar{\bf W}, {\bf Y} } \, 
	\tr \Big\lbrace \bar{\bf W} \big[ \big( {\bf Y} {\bf X} {\bf C} \! - \! {\bf I} \big) \boldsymbol{\Phi}^{-1} \big( {\bf Y} {\bf X} {\bf C} \! - \! {\bf I} \big)^\hem \! + \! {\bf Y} \boldsymbol{\Pi}^{-1} {\bf Y}^\hem \big] \Big\rbrace 
	\! - \! \log \det \big( \bar{\bf W} \big),
	\ \,
	{\rm s.t.} \, {\bf X} \in \mathcal{X},
	\label{Opt_Eq_WMSE}
\end{align}
where $ \bar{\bf W} \! = \! \boldsymbol{\Phi} + {\bf C}^\hem {\bf X}^\hem \boldsymbol{\Pi} {\bf X} {\bf C} $ is a positive semi-definite weighting matrix.

\noindent\textit{Proof}: The proof is relegated to \textbf{Appendix~\ref{sec:appendix-D}}.

Note that given the weighting matrix $ \bar{\bf W} $, the optimization problem (\ref{Opt_Eq_WMSE}) shares the similar form as (\ref{Opt_2}). 
Based on Proposition~3 and the weighted MSE performance metric (\ref{Obj_WMSE}), we introduce a unified optimization objective to unify these performance metrics, and the transceiver optimization is formulated as
\begin{subequations}
	\label{Opt_2}
	\begin{align}
		\min_{ {\bf F} } \ \, & \tr \big\lbrace {\bf A}^{\!\hem} \big[ ( {\bf G} {\bf H}_{\rm c} {\bf F} {\bf C} - {\bf I} ) ( {\bf G} {\bf H}_{\rm c} {\bf F} {\bf C} - {\bf I} )^\hem + {\bf G} {\bf R}_{{\bf n}_{\rm c}} {\bf G}^\hem \big] {\bf A} \big\rbrace \\
		{\rm s.t.} \ \ & \tr( {\bf F} {\bf F}^\hem ) = P_{\rm max}, 
		\label{Opt_2:C1} \\
		& \tr \big( {\bf F}^\hem {\bf B} {\bf F} \big) \ge \bar{\gamma}_{\rm D},
		\label{Opt_2:C2} 
	\end{align}
\end{subequations}
where $ {\bf A} $ and $ {\bf C} $ are weighting matrices. Therefore, we focus our attention on the unified MSE minimization problem (\ref{Opt_2}) to demonstrate the proposed optimization framework.
It is seen that the two constraints are coupled in optimization problem (\ref{Opt_2}) which hinders the optimization of the BS precoder. Fortunately, by referring to the maximum power allocation criterion derived from Proposition~2, the sum power constraint can be eliminated, and the problem (\ref{Opt_2}) can be equivalently transformed into \cite{QingjiangShi2023Rethinking}
\begin{align}
	\min_{ {\bf F} } \ \, & \tr \bigg\lbrace {\bf A}^{\!\hem} \bigg[ ( {\bf G} {\bf H}_{\rm c} \widehat{\bf F} {\bf C} - {\bf I} ) ( {\bf G} {\bf H}_{\rm c} \widehat{\bf F} {\bf C} - {\bf I} )^\hem 
	+ {\bf G} {\bf R}_{{\bf n}_{\rm c}} {\bf G}^\hem \frac{ \tr \big( \widehat{\bf F}^\hem \widehat{\bf F} \big) }{ P_{\rm max} } \bigg] {\bf A} \bigg\rbrace \notag \\
	{\rm s.t.} \ \, & \tr \big( \widehat{\bf F}^\hem \widehat{\bf B} \widehat{\bf F} \big) \le 0,
	\label{Opt_3}
\end{align}
where the matrix $ \widehat{\bf B} $ is defined as $ \widehat{\bf B} = \frac{\bar{\gamma}_{\rm D}}{P_{\rm max}} {\bf I} - {\bf B} $. Note that the optimization problem (\ref{Opt_3}) is a standard quadratically constrained quadratic programing (QCQP) problem, and the optimal precoder $ \widehat{\bf F}_{\rm opt} $ can be found using the first-order derivative, i.e.,
\begin{align}
	\vecm(\widehat{\bf F}_{\rm opt}) = \bigg( ( {\bf C}^* {\bf C}^\T ) \otimes ( {\bf H}^\hem {\bf G}^\hem {\bf G} {\bf H} ) + {\bf I} \otimes \Big( \frac{\tr( {\bf G} {\bf R}_{{\bf n}_{\rm c}} {\bf G}^\hem )}{P_{\rm max}} + \mu \widehat{\bf B} \Big) \bigg)^{\!\!-1} \! \vecm\big( {\bf H}^\hem {\bf G}^\hem {\bf C}^\hem \big)
\end{align}
in which the scalar $ \mu $ is a Lagrange multiplier. It is worth noting that the matrix $ \widehat{\bf B} $ is neither semi-positive definite nor semi-negative definite. Henceforth, the left-hand side of the radar constraint no longer monotonically varies with the multiplier $ \mu $. Therefore, the subgradient algorithm is proposed to search for the optimal $ \mu $\cite{Boyd2003Subgradient}.

\vspace{-5.5mm}

\subsection{ADMM-Based RIS Phase Shift Optimization}
\label{sec:single_ISAC:RIS}

\vspace{-1.5mm}

With the BS precoder $ {\bf F} $, the radar receive beamformer $ {\bf w} $, and the user equalizer $ {\bf G} $ optimized, the remaining task of the ISAC system design is to adjust the RIS phase shifts to minimize the communication objective while keeping the detection performance at the same time. The RIS phase shift design is thus formulated as
\begin{subequations}
	\label{RIS_Opt_0}
	\begin{align}
		\min_{ \boldsymbol{\Theta} } \ & f_{\rm obj} \big( \boldsymbol{\Phi}_{\rm MSE} \big) \\
		{\rm s.t.} \ \ & 
		\tr( {\bf F}^\hem {\bf B} {\bf F} ) \ge \bar{\gamma}_{\rm D}, \label{RIS_Opt_0:C1} \\
		& | \theta_i | = 1, \ \ i = 1, 2, \ldots, N_{\rm R}. \label{RIS_Opt_0:C2}
	\end{align}
\end{subequations}
To the best of the authors' knowledge, the RIS phase shift design (\ref{RIS_Opt_0}) is extremely difficult to be optimized, and the major challenges come from the coupling nature of the two non-convex constraints and the quartic property of the radar constraint w.r.t. the RIS phase shifts. 

Aiming at the removal of the radar constraint's obstacle, we define the following auxiliary matrices
\begin{align}
	& {\bf X} \triangleq \big| \rho_{\rm tg} \big| \Diag( {\bf w}^\T {\bf H}_{\rm BR}^* ) {\bf a}_{\rm RIS,r}^* {\bf a}_{\rm RIS,r}^\T \Diag( {\bf H}_{\rm BR}^\T {\bf w}^* ), \notag \\
	& {\bf Y} \triangleq \big| \rho_{\rm tg} \big| \Diag( {\bf a}_{\rm RIS,t}^\T ) {\bf H}_{\rm BR}^\T {\bf F}^* {\bf F}^\T {\bf H}_{\rm BR}^* \Diag( {\bf a}_{\rm RIS,t}^* ),
	\label{Def_RIS_X_Y_SU}
\end{align}
and $ \boldsymbol{\theta} \triangleq \diag \{ \boldsymbol{\Theta} \} $. The radar constraint (\ref{RIS_Opt_0:C1}) can thus be rewritten as
\begin{align}
	\tr ( \boldsymbol{\theta}^\hem {\bf X} \boldsymbol{\theta} \cdot \boldsymbol{\theta}^\hem {\bf Y} \boldsymbol{\theta} )
	\ge \bar{\gamma}_{\rm D},
	\label{Radar_RIS_Quartic}
\end{align}
which is a non-convex quartic constraint. Then, we embark on the majorization-minimization (MM) method to find a proper convex approximation of (\ref{Radar_RIS_Quartic}) and optimize the relaxed problem iteratively. Let $ \boldsymbol{\Omega} \triangleq \boldsymbol{\theta} \boldsymbol{\theta}^\hem $, the quartic radar constraint (\ref{Radar_RIS_Quartic}) is transformed into
\begin{align}
	\tr ( {\bf X} \boldsymbol{\Omega} {\bf Y} \boldsymbol{\Omega} ) 
	& \overset{(a)}{\ge} \vecm^\hem\!( \boldsymbol{\Omega}_t ) \big( {\bf X}^\T \otimes {\bf Y} \big) \vecm ( \boldsymbol{\Omega} )
	+ \vecm^\hem\!( \boldsymbol{\Omega} ) \big( {\bf X}^\T \otimes {\bf Y} \big) \vecm ( \boldsymbol{\Omega}_t )
	\notag \\ & 
	~~~~~- \vecm^\hem\!( \boldsymbol{\Omega}_t ) \big( {\bf X}^\T \otimes {\bf Y} \big) \vecm ( \boldsymbol{\Omega}_t ) 
\notag \\ &
\overset{(b)}{=} 
\boldsymbol{\theta}^\hem {\bf Q} \boldsymbol{\theta} - \tr( {\bf X} \boldsymbol{\theta}_t \boldsymbol{\theta}_t^\hem {\bf Y} \boldsymbol{\theta}_t \boldsymbol{\theta}_t^\hem ) \ge \bar{\gamma}_{\rm D},
	\label{Radar_Constraint_1}
\end{align}
where $ {\bf Q} = {\bf X} \boldsymbol{\theta}_t \boldsymbol{\theta}_t^\hem {\bf Y} + {\bf Y}^\hem \boldsymbol{\theta}_t \boldsymbol{\theta}_t^\hem {\bf X}^\hem $ and $ \boldsymbol{\theta}_t $ denotes the RIS phase shifts at $ t $th iteration. The inequality (a) is derived by vectorizing the matrix $ \boldsymbol{\Omega} $ and using the first-order Taylor expansion, and the equation (b) is obtained by recovering the vectorized matrix $ \boldsymbol{\Omega} $.
Since the constraint (\ref{Radar_Constraint_1}) is still non-convex, we further define the auxiliary variables $ \widehat{\bf Q} \! \triangleq \! \lambda_{\rm max}( {\bf Q} ) {\bf I} - {\bf Q} $ and $ c \triangleq N_{\rm R} \lambda_{\rm max}( {\bf Q} ) - \tr( \boldsymbol{\theta}_t^\hem {\bf Y} \boldsymbol{\theta}_t \boldsymbol{\theta}_t^\hem {\bf X} \boldsymbol{\theta}_t ) - \bar{\gamma}_{\rm D} $ and express the radar constraint (\ref{Radar_Constraint_1}) into a convex form
\begin{align}
	\boldsymbol{\theta}^\hem \widehat{\bf Q} \boldsymbol{\theta} \le c.
	\label{Radar_Constraint_cvx}
\end{align}

After transforming the quartic radar constraint (\ref{RIS_Opt_0:C1}) into the convex constraint (\ref{Radar_Constraint_cvx}), the primary challenge for the RIS phase shift design is to decouple the constraints. Therefore, we resort to iterative methods and propose an ADMM-based RIS phase shift design \cite{boyd2011ADMM}, where an auxiliary variable $ \boldsymbol{\alpha} $ is introduced to decouple the radar constraint and the non-convex unit-modulus constraints. Then, the RIS phase shift design can be written as
\begin{subequations}
	\label{RIS_Opt_ADMM_1}
	\begin{align}
		\min_{ \boldsymbol{\theta}, \boldsymbol{\alpha} } \ & f_{\rm obj} \big( \boldsymbol{\Phi}_{\rm MSE} \big) \\
		{\rm s.t.} \ \ 
		& \boldsymbol{\theta}^\hem \widehat{\bf Q} \boldsymbol{\theta} \le c, \label{RIS_Opt_ADMM_1:C1} \\
		& \boldsymbol{\alpha} = \boldsymbol{\theta}, \label{RIS_Opt_ADMM_1:C2} \\
		& | \alpha_i | = 1, \ \ i = 1, 2, \ldots, N_{\rm R}. \label{RIS_Opt_ADMM_1:C3}
	\end{align}
\end{subequations}
By adopting the ADMM framework, the equality constraint (\ref{RIS_Opt_ADMM_1:C2}) is attached to the objective function, and the scaled augmented Lagrange function is expressed as
\begin{align}
	\mathcal{L}_{\rm RIS} ( \boldsymbol{\theta}, \boldsymbol{\alpha}, \boldsymbol{\upsilon} ) = f_{\rm obj} \big( \boldsymbol{\Phi}_{\rm MSE} \big) + \frac{\varrho}{2} \big\Vert \boldsymbol{\alpha} - \boldsymbol{\theta} + \frac{\boldsymbol{\upsilon}}{\varrho} \big\Vert_2^2,
\end{align}
where $ \boldsymbol{\upsilon} $ is a dual variable and $ \varrho $ is a penalty parameter. 
Then, following the methodology of ADMM, the iterative updates of variables are given by
\begin{subequations}
	\begin{align}
		& \boldsymbol{\alpha}^{(t+1)} := \arg\min_{\boldsymbol{\alpha}} \mathcal{L}_{\rm RIS} ( \boldsymbol{\theta}^{(t)}, \boldsymbol{\alpha}, \boldsymbol{\upsilon}^{(t)} ), \\
		& \boldsymbol{\theta}^{(t+1)} := \arg\min_{\boldsymbol{\theta}} \mathcal{L}_{\rm RIS} ( \boldsymbol{\theta}, \boldsymbol{\alpha}^{(t+1)}, \boldsymbol{\upsilon}^{(t)} ), \\
		& \boldsymbol{\upsilon}^{(t+1)} := \boldsymbol{\upsilon}^{(t)} + ( \boldsymbol{\alpha}^{(t+1)} - \boldsymbol{\theta}^{(t+1)} ).
		\label{Update_upsilon_SU}
	\end{align}
\end{subequations}

Firstly, considering that the auxiliary variable $ \boldsymbol{\alpha} $ appears only in the penalty term and the unit-modulus constraints, its optimal solution can be directly derived as
\begin{align}
	\boldsymbol{\alpha}_{\rm opt} = e^{j\measuredangle ( \boldsymbol{\theta} - \boldsymbol{\upsilon} )},
	\label{Optimal_Alpha_SU}
\end{align}
where $ \measuredangle {\bf a} $ denotes the angle vector of a complex vector $ {\bf a} $. Then, we move our attention to the RIS phase shift optimization. It is also worth noting that different performance metrics would lead to the distinct RIS phase shifts. 

In order to derive a general optimization framework, the unified objective function in Section~\ref{sec:single_ISAC:BS} is adopted as well and the RIS phase shift optimization in (\ref{RIS_Opt_ADMM_1}) can also be transformed into the equivalent MSE minimization problem, which is given by
\begin{align}
	\min_{ \boldsymbol{\theta} } \ 
	& \tr \big\lbrace {\bf A}^{\!\hem} \big[ ( {\bf G} {\bf H}_{\rm c} {\bf F} {\bf C} - {\bf I} ) ( {\bf G} {\bf H}_{\rm c} {\bf F} {\bf C} - {\bf I} )^\hem + {\bf G} {\bf R}_{{\bf n}_{\rm c}} {\bf G}^\hem \big] {\bf A} \big\rbrace 
	+ \frac{\varrho}{2} \big\Vert \boldsymbol{\alpha} - \boldsymbol{\theta} + \boldsymbol{\upsilon} \big\Vert_2^2 \notag \\
	{\rm s.t.} \ \,
	& \boldsymbol{\theta}^\hem \widehat{\bf Q} \boldsymbol{\theta} \le c.
	\label{RIS_ADMM_WMSE_1}
\end{align}
Since the RIS phase shifts are implicitly included in $ {\bf H}_{\rm c} $, we can rewrite the objective function as
\begin{align}
	\min_{ \boldsymbol{\Theta}, \boldsymbol{\theta} } \ \ 
	& \tr \Big\lbrace \bar{\bf H}_{\rm RU} \boldsymbol{\Theta} \bar{\bf H}_{\rm BR}^\hem \bar{\bf H}_{\rm BR} \boldsymbol{\Theta}^\hem \bar{\bf H}_{\rm RU}^\hem
	+ \bar{\bf H}_{\rm RU} \boldsymbol{\Theta} \bar{\bf H}_{\rm BR}^\hem ( {\bf A}^{\!\hem} {\bf G} {\bf H}_{\rm BU} {\bf F} {\bf C} - {\bf A}^{\!\hem} )^\hem \notag \\
	& \quad
	+ ( {\bf A}^{\!\hem} {\bf G} {\bf H}_{\rm BU} {\bf F} {\bf C} - {\bf A}^{\!\hem} ) \bar{\bf H}_{\rm BR} \boldsymbol{\Theta}^\hem \bar{\bf H}_{\rm RU}^\hem \Big\rbrace
	+ \frac{\varrho}{2} \big\Vert \boldsymbol{\alpha} - \boldsymbol{\theta} + \boldsymbol{\upsilon} \big\Vert_2^2,
	\label{RIS_ADMM_WMSE_2}
\end{align}
where the matrices $ \bar{\bf H}_{\rm RU} $ and $ \bar{\bf H}_{\rm BR} $ are defined as
\begin{align}
	& \bar{\bf H}_{\rm RU} \! = \! \big[ \bar{\bf h}_{{\rm RU},1}, \ldots, \bar{\bf h}_{{\rm RU},N_{\rm R}} \big] \! \triangleq \! {\bf A}^{\!\hem} {\bf G} {\bf H}_{\rm RU}, 
	\ \ 
	\bar{\bf H}_{\rm BR} \! = \! \big[ \bar{\bf h}_{{\rm BR},1}, \ldots, \bar{\bf h}_{{\rm BR},N_{\rm R}} \big] \! \triangleq \! {\bf C}^\hem {\bf F}^\hem {\bf H}_{\rm BR},
\end{align}
for notational simplicity.
To unify the representation of RIS phase shifts, we again rewrite the objective function \eqref{RIS_ADMM_WMSE_2} as a quadratic function of the RIS phase shifts $ \boldsymbol{\theta} \triangleq \diag \{ \boldsymbol{\Theta} \} $, i.e.,
\begin{align}
	\min_{ \boldsymbol{\theta} } \ \, & \boldsymbol{\theta}^\hem \bigg( {\bf H}_{\boldsymbol{\Theta}}^\hem {\bf H}_{\boldsymbol{\Theta}} + \frac{\varrho}{2} {\bf I} \bigg) \boldsymbol{\theta}
	+ 2 \Re \bigg\lbrace \! \bigg( {\bf H}_{\boldsymbol{\Theta}}^\hem \vecm \big( {\bf A}^{\!\hem} {\bf G} {\bf H}_{\rm BU} {\bf F} {\bf C} \! - \! {\bf A}^{\!\hem} \big) \! - \! \varrho \Big( \frac{\boldsymbol{\alpha}}{2} \! + \! \frac{\boldsymbol{\upsilon}}{2} \Big) \bigg)^\hem \boldsymbol{\theta} \bigg\rbrace \notag \\
	{\rm s.t.} \ \ 
	& \boldsymbol{\theta}^\hem \widehat{\bf Q} \boldsymbol{\theta} \le c,
	\label{RIS_ADMM_theta_2}
\end{align}
where the auxiliary matrix $ {\bf H}_{\boldsymbol{\Theta}} $ is defined as $ {\bf H}_{\boldsymbol{\Theta}} \triangleq \big[ \vecm ( \bar{\bf h}_{{\rm RU},1} \bar{\bf h}_{{\rm BR},1}^\hem ), \ldots, \vecm ( \bar{\bf h}_{{\rm RU},N_{\rm R}} \bar{\bf h}_{{\rm BR},N_{\rm R}}^\hem ) \big] $. Note that the optimization problem (\ref{RIS_ADMM_theta_2}) is a standard QCQP problem with only a single quadratic constraint \cite{Boyd04}. 
Then, the optimal RIS phase shifts $ \boldsymbol{\theta} $ to the problem (\ref{RIS_ADMM_theta_2}) is derived as
\begin{align}
	\boldsymbol{\theta}_{\rm opt} = - \bigg( {\bf H}_{\boldsymbol{\Theta}}^\hem {\bf H}_{\boldsymbol{\Theta}} + \frac{\varrho}{2} {\bf I} + \delta \widehat{\bf Q} \bigg)^{\!\!-1}
	{\bf g},
	\label{Optimal_theta_SU}
\end{align}
where $ \delta $ is the Lagrange multiplier and $ {\bf g} \triangleq {\bf H}_{\boldsymbol{\Theta}}^\hem \vecm \big( {\bf A}^{\!\hem} {\bf G} {\bf H}_{\rm BU} {\bf F} {\bf C} - {\bf A}^{\!\hem} \big) - \varrho \big( \frac{\boldsymbol{\alpha}}{2} + \frac{\boldsymbol{\upsilon}}{2} \big) $. 
Since the dual problem of the optimization problem (\ref{RIS_ADMM_theta_2}) belongs to scalar optimization, its derivative is monotonically increasing w.r.t. the dual variable $ \delta $, the optimal $ \delta $ and $ \boldsymbol{\theta} $ can thus be found by the bisection method.

Therefore, the locally optimal solutions of RIS phase shifts $ \boldsymbol{\Theta} $ are gradually obtained through an iterative manner. For the reader's convenience, the complete procedure of the RIS optimization is summarized in Algorithm~\ref{alg_admm}.

\begin{algorithm}[!t]
	\caption{The RIS Phase Shift Design Algorithm}
	\label{alg_admm}
	\begin{algorithmic}[1]
		\REQUIRE{The BS precoder $ {\bf F} $, the user equalizer $ {\bf G} $, the radar receiver $ {\bf w} $, the channel matrices $ {\bf H}_{\rm BR}, {\bf H}_{\rm RU}, {\bf H}_{\rm BU}, {\bf a}_{{\rm RIS}, t}, {\bf a}_{{\rm RIS}, r} $, the reflection efficient $ \rho_{\rm tg} $, the effective radar detection threshold $ \bar{\gamma}_{\rm D} $, the penalty parameter $ \varrho $ and the convergence threshold $ \varepsilon $. }
		\STATE Randomly initialize RIS phase shift $ \boldsymbol{\theta}^{(0)} $ and set the iteration number $ t = 0 $;
		\STATE Compute auxiliary matrices $ {\bf X}, {\bf Y} $ and normalized detection threshold $ \bar{\gamma}_{\rm D} $;
		\REPEAT
		\STATE Initialize $ \boldsymbol{\alpha} = \boldsymbol{\theta}^{(t)} $, $ \boldsymbol{\upsilon} = {\bf 0} $, and the ADMM iteration number $ k = 0 $;
		\STATE Calculate matrix $ \widehat{\bf Q}^{(t)} $ and $ c^{(t)} $;
		\REPEAT
		\STATE Compute the effective channel matrices $ \bar{\bf H}_{\rm RU}, \bar{\bf H}_{\rm BR}, {\bf H}_{\boldsymbol{\Theta}}^{(t)} $ and the vector $ {\bf g}^{(t)} $;
		\STATE Calculate the optimal $ \boldsymbol{\alpha} $ based on (\ref{Optimal_Alpha_SU});
		\STATE Compute the optimal $ \boldsymbol{\theta}^{(t+1)}_{(k)} $ and the multiplier $ \delta $ according to (\ref{Optimal_theta_SU});
		\STATE Update dual variable $ \boldsymbol{\upsilon} $ using (\ref{Update_upsilon_SU}) and $ k = k + 1 $;
		\UNTIL{ $ \Vert \boldsymbol{\alpha} - \boldsymbol{\theta}^{(t+1)}_{(k)} \Vert_2 \le \varepsilon $; }
		\STATE Update iteration number $ t = t + 1 $;
		\UNTIL{ $ \Vert \boldsymbol{\theta}^{(t+1)} - \boldsymbol{\theta}^{(t)} \Vert_2 \le \varepsilon $; }
		\RETURN{The optimal RIS phase shifts $ \boldsymbol{\Theta}^{(t)} = {\bf diag} ( \boldsymbol{\theta}^{(t)} ) $.}
	\end{algorithmic}
\end{algorithm}

\vspace{-6mm}

\subsection{Computational Efficient RIS Design Algorithm}
\label{sec:low_complex_alg}

\vspace{-1.5mm}

In the above subsection, we propose an ADMM-based RIS design to reach the performance limitation of the ISAC system. However, the multi-layer iterations of Algorithm~\ref{alg_admm} contribute to a high computational complexity, which is not suitable for future implementation. To alleviate this issue, we develop a low-complexity RIS shifts design in this subsection.

In general, the RIS is responsible for balancing the transmit beams between the user and the target. To achieve favorable sensing function, it is reasonable to design the RIS phase shifts by sacrificing some communication performance in exchange for lower complexity. Moreover, by pointing the RIS beam at the radar target, the feasible set of the BS precoder is also enlarged, and thus provide extra freedom of precoder design. Based on this philosophy, the RIS optimization problem is reformulated as
\begin{align}
	\min_{ \boldsymbol{\theta} } \ & - \tr ( \boldsymbol{\theta}^\hem {\bf X} \boldsymbol{\theta} \!\cdot\! \boldsymbol{\theta}^\hem {\bf Y} \boldsymbol{\theta} ) \notag \\
	{\rm s.t.} \ \,
	& \, | \theta_i | = 1, \ \, i = 1, 2, \ldots, N_{\rm R}.
	\label{Low Complex_RIS_0}
\end{align}
The optimization problem (\ref{Low Complex_RIS_0}) is clearly non-convex. By introducing a smoothly varying inner product $ \langle \theta_1, \theta_2 \rangle = \Re \{ \theta_1^* \theta_2 \} $ and the corresponding Riemannian manifold \cite{ManifoldsOptimization2009Absil,ZhangJun2016Rieman}
\begin{align}
	\mathcal{M}_{\rm RIS} = \{ \boldsymbol{\theta} \in \mathbb{C}^{N_{\rm R} \times 1}: \langle \theta_1, \theta_1 \rangle = \cdots = \langle \theta_{N_{\rm R}}, \theta_{N_{\rm R}} \rangle = 1 \},
	\label{Def_RiemanianManifold}
\end{align}
the unit-modulus constrained RIS design (\ref{Low Complex_RIS_0}) can be optimized using the steepest descend method \cite{ZhangJun2016Rieman}.
Based on (\ref{Def_RiemanianManifold}), tangent vectors are introduced to describe the steepest moving direction of curves in the manifold. The tangent space at a point $ \boldsymbol{\theta} $ on the manifold $ \mathcal{M}_{\rm RIS} $ is described as the set of all possible tangent vectors at $ \boldsymbol{\theta} $ on $ \mathcal{M}_{\rm RIS} $, i.e.,
$ \mathcal{T}_{\boldsymbol{\theta}} \mathcal{M}_{\rm RIS} \!=\! \big\{ \boldsymbol{\delta} \in \mathbb{C}^{N_{\rm R} \times 1}: \Re \{ \boldsymbol{\delta} \circ \boldsymbol{\theta}^* \} = {\bf 0} \big\} $.
Armed with the tangent space definition, we formulate the Riemannian gradient of the smooth objective function $ \ell( \boldsymbol{\theta} ) = - \tr ( \boldsymbol{\theta}^\hem {\bf X} \boldsymbol{\theta} \!\cdot\! \boldsymbol{\theta}^\hem {\bf Y} \boldsymbol{\theta} ) $ at the point $ \boldsymbol{\theta} $ as
\begin{align}
	\nabla_{\rm R} \ell = \nabla \ell - \Re \{ \nabla^* \ell \circ \boldsymbol{\theta} \} \circ \boldsymbol{\theta},
	\label{Eq_Riemannian_Gradient}
\end{align}
where $ \nabla \ell $ denotes the Euclidean gradient of the function $ \ell( \cdot ) $ w.r.t. the RIS phase shifts $ \boldsymbol{\theta} $, i.e.,
\begin{align}
	\nabla \ell = - \boldsymbol{\theta}^\hem {\bf X} \boldsymbol{\theta} {\bf Y} \boldsymbol{\theta} - \boldsymbol{\theta}^\hem {\bf Y} \boldsymbol{\theta} {\bf X} \boldsymbol{\theta}.
	\label{Eq_Euclidean_Gradient}
\end{align}

\begin{algorithm}[!t]
	\caption{The Conjugate Gradient based Computation Efficient Algorithm}
	\label{alg_cgd}
	\begin{algorithmic}[1]
		\REQUIRE{The auxiliary matrices $ {\bf X}, {\bf Y} $ and the convergence threshold $ \varepsilon $. }
		\STATE Randomly initialize RIS phase shift $ \boldsymbol{\theta}^{(0)} $ and set the iteration number $ t = 0 $;
		\STATE Compute the Riemannian gradient $ \nabla_{\rm R} \ell^{(0)} $ and initialize conjugate gradient as $ \nabla_{\bf cg} \ell^{(0)} $;
		\REPEAT
		\STATE Compute gradient step size $ \zeta^{(t)} $ based on line search method;
		\STATE Calculate the retracted phase shifts $ \boldsymbol{\theta}^{(t+1)} = \mathcal{R} \big( \boldsymbol{\theta}^{(t)} + \zeta^{(t)} \nabla_{\bf cg} \ell^{(t)} \big) $;
		\STATE Compute the Euclidean gradient of the objective $ \nabla \ell^{(t+1)} $ based on (\ref{Eq_Euclidean_Gradient}), then update the Riemannian gradient $ \nabla_{\rm R} \ell^{(t+1)} $ based on (\ref{Eq_Riemannian_Gradient});
		\STATE Calculate the Polak-Ribiere parameter by $ \beta^{(t+1)} = \frac{ \nabla_{\rm R}^\hem \ell^{(t+1)} \big[ \nabla_{\rm R} \ell^{(t+1)} 
			- \mathcal{T}_{(t) \rightarrow (t+1)} ( \nabla_{\rm R} \ell^{(t)} ) \big] }
		{ \Vert \mathcal{T}_{(t) \rightarrow (t+1)}( \nabla_{\rm R} \ell^{(t)} ) \Vert_2^2 } $
		\STATE Update the conjugate gradient by $ \nabla_{\bf cg} \ell^{(t+1)} = - \nabla_{\rm R} \ell^{(t+1)} + \beta^{(t+1)} \mathcal{T}_{(t) \rightarrow (t+1)} ( \nabla_{\bf cg} \ell^{(t)} ) $;
		\STATE Update iteration number $ t = t + 1 $;
		\UNTIL{ $ \Vert \boldsymbol{\theta}^{(t)} - \boldsymbol{\theta}^{(t-1)} \Vert_2 \le \varepsilon $; }
		\RETURN{The optimal RIS phase shifts $ \boldsymbol{\Theta}^{(t)} = {\bf diag} ( \boldsymbol{\theta}^{(t)} ) $.}
	\end{algorithmic}
\end{algorithm}
Before executing the gradient descent method, it is noticed that the tangent space and the Riemannian manifold $ \mathcal{M}_{\rm RIS} $ may not be consistent, which makes that the phase shifts $ \boldsymbol{\theta} $ no longer locate in the same manifold after each update. Thus, a retraction mapping $ \mathcal{R}: \mathcal{T}_{\boldsymbol{\theta}} \mathcal{M}_{\rm RIS} \rightarrow \mathcal{M}_{\rm RIS} $ is introduced to force the phase shifts in the Riemannian manifold during iterations
\begin{align}
	\mathcal{R} ( \boldsymbol{\alpha} ) = e^{ j \measuredangle \boldsymbol{\alpha} }.
\end{align}
On the other hand, a vector transport $ \mathcal{T}_{(t) \rightarrow (t+1)} (\cdot): \mathcal{T}_{\boldsymbol{\theta}^{(t)}} \mathcal{M}_{\rm RIS} \rightarrow \mathcal{T}_{\boldsymbol{\theta}^{(t+1)}} \mathcal{M}_{\rm RIS} $ is proposed so that the tangent vector in the previous iteration can be mapped into the tangent space of the current iteration, which is given by 
\begin{align}
	\mathcal{T}_{(t) \rightarrow (t+1)} ( {\bf z}( \boldsymbol{\theta}^{(t)} ) ) = {\bf z}( \boldsymbol{\theta}^{(t)} ) - \Re\big\{ {\bf z}^*( \boldsymbol{\theta}^{(t)} ) \circ \boldsymbol{\theta}^{(t+1)} \big\} \circ \boldsymbol{\theta}^{(t+1)}.
\end{align}
Henceforth, the conjugate gradient algorithm can be utilized considering its super-linear convergence advantage, where the conjugate gradient is updated according to $ \nabla_{\bf cg} f^{(t+1)} = \nabla f^{(t+1)} + \beta^{(t+1)} \nabla_{\bf cg} f^{(t)} $, with $ \beta^{(t+1)} $ being the Polak-Ribiere parameter \cite{ZhangJun2016Rieman}.
The optimal RIS phase shifts can thus be efficiently calculated and the detailed algorithm is presented in Algorithm~\ref{alg_cgd}.

\vspace{-3mm}

\section{ISAC System Optimization For the MUMT Scenarios}
\label{sec:multi_ISAC}

\vspace{-1.5mm}

In Section~\ref{sec:single_ISAC}, we investigate the ISAC system design for the SUST scenario. 
Based on the proposed design, the DFRC-BS is able to sense multiple targets by traversing all possible directions. 
However, for the slow-moving target sensing cases, where the targets' prior information can be obtained by the BS based on the previous detection, the traversal scheme would be less economical. This motivates us to investigate an efficient MTMU ISAC system design.

\vspace{-6mm}

\subsection{Multiple-Target Radar Receiver Design}

\vspace{-1.5mm}

According to the problem formulation in (\ref{Radar_Beamformer_Opt_Prob}), it is demonstrated that the interference signals could severely deteriorate the detection performance, and an interference-free strategy is adopted to show the performance limitation for an ISAC transceiver. To achieve this goal, the ZF receiver is utilized. To be specific, define the correlation matrix of interferences of target $ k $
\begin{align}
	\boldsymbol{\Upsilon}_k 
	= \, & {\bf H}_{\rm BR} \boldsymbol{\Theta} 
	[ {\bf A}_{\rm RIS,r} ]_{:, -k} [ \boldsymbol{\Xi} ]_{-k,-k} [ {\bf A}_{\rm RIS,t} ]_{:,-k}^\hem \boldsymbol{\Theta} {\bf H}_{\rm BR}^\hem {\bf F} \notag \\
	& \times {\bf F}^\hem {\bf H}_{\rm BR} \boldsymbol{\Theta}^\hem
	[ {\bf A}_{\rm RIS,t} ]_{:,-k} [ \boldsymbol{\Xi} ]_{-k,-k}^\hem [ {\bf A}_{\rm RIS,r} ]_{:, -k}^\hem {\bf H}_{\rm BR}^\hem,
\end{align}
and the optimal $ k $th radar beamformer $ {\bf w}_k $ should lie in the zero space $ \mathcal{Z} ( \boldsymbol{\Upsilon}_k ) $, i.e.,
\begin{align}
	{\bf w}_k^{\rm opt} = \arg \max \, \big\Vert \rho_{{\rm tg},k} {\bf w}_k^\hem {\bf H}_{\rm BR} \boldsymbol{\Theta} {\bf a}_{{\rm RIS,r},k} {\bf a}_{{\rm RIS,t},k}^\hem \boldsymbol{\Theta} {\bf H}_{\rm BR}^\hem {\bf F} \big\Vert^2,
	\ \ 
	{\bf w}_k \in \mathcal{Z} ( \boldsymbol{\Upsilon}_k ).
\end{align}
Substituting it into (\ref{Eq_Detected_Sig_MT}), the target detection tasks are decoupled and can be independently executed using the GLRT defined in Section~\ref{sec:sys_model:problem}.
Further, by referring to Proposition~1, the radar constraints in the multiple targets scenarios are transformed into 
\begin{align}
	\tr \big( {\bf F}^\hem {\bf B}_{\ell} {\bf F} \big) \ge \bar{\gamma}_{{\rm D},\ell},
	\ \ \ell = 1, \ldots, L,
	\label{Radar_Constraint_MT}
\end{align}
where 
$ {\bf B}_{\ell} = | \rho_{{\rm tg},\ell} |^2 {\bf H}_{\rm BR}^\hem \boldsymbol{\Theta}^\hem {\bf a}_{{\rm RIS,t},\ell} {\bf a}_{{\rm RIS,r},\ell}^\hem \boldsymbol{\Theta}^\hem {\bf H}_{\rm BR} {\bf w}_{\ell} {\bf w}_{\ell}^\hem {\bf H}_{\rm BR}^\hem \boldsymbol{\Theta} {\bf a}_{{\rm RIS,r},\ell} {\bf a}_{{\rm RIS,t},\ell}^\hem \boldsymbol{\Theta} {\bf H}_{\rm BR} $ and $ \bar{\gamma}_{{\rm D},\ell} = \frac{ \sigma_{n,r}^2 \Vert {\bf w}_{\ell} \Vert_2^2 \bar{P}^{-1} ( \gamma_{\rm D} ) }{ D } $.

\vspace{-5.5mm}

\subsection{Optimal ISAC Transceiver Design}

\vspace{-1.5mm}

Compared with problem (\ref{Opt_1}) in the SUST scenario, the transceiver design (\ref{General_Opt_Model}) in the MUMT scenario mainly originates from the communication interference and the additional radar constraints. Following the philosophy in Section~\ref{sec:single_ISAC}, we take the equivalent weighted MSE minimization as an example to extend our proposed design to the MUST case. Thus, the transceiver design is written as
\begin{subequations}
	\label{Opt_MUMT_0}
	\begin{align}
		\min_{\substack{ {\bf F}, \boldsymbol{\Theta}, \{ {\bf G}_k \} } } \ 
		& \sum\nolimits_{k = 1}^K \tr \big\lbrace {\bf A}_k^\hem \boldsymbol{\Phi}_{{\rm MSE},k} {\bf A}_k \big\rbrace \\
		{\rm s.t.} \ \ \ & \tr( {\bf F} {\bf F}^\hem ) \le P_{\rm max}, \label{Opt_MUMT_0:C1} \\
		& \tr \big( {\bf F}^\hem {\bf B}_{\ell} {\bf F} \big) \ge \bar{\gamma}_{{\rm D},\ell}, \ \ell = 1, \ldots, L.
		\label{Opt_MUMT_0:C2}
	\end{align}
\end{subequations}

Noticing that there is no constraint on the equalizers $ \{ {\bf G}_k \} $, henceforth, the optimal equalizers can be directly obtained using the LMMSE criterion\cite{Xin2022TCOM}, i.e.,
\begin{align}
	{\bf G}_{{\rm opt},k} = {\bf F}_k^\hem {\bf H}_{{\rm c},k}^\hem ( {\bf H}_{{\rm c},k} {\bf F} {\bf F}^\hem {\bf H}_{{\rm c},k}^\hem + {\bf R}_{{\bf n}_{\rm c}} )^{-1}.
\end{align}

However, the DFRC-BS precoder optimization is not quite simple, since there are several constraints on the precoder, and the radar constraints (\ref{Opt_MUMT_0:C1}) are non-convex. Thus, we manage to reduce the number of the constraints then resort to the MM method once again to relax the non-convex constraints to maneuverable linear ones. Firstly, it is worth highlighting that the maximum power allocation scheme is also valid in the MUMT scenario, which can also be proved as in \cite{QingjiangShi2023Rethinking}. Thus, by replacing the BS precoder $ {\bf F}_k $ with the scaled precoder $ \widehat{\bf F}_k \triangleq \frac{\tr(\widehat{\bf F}_k \widehat{\bf F}_k^\hem)}{P_{\rm max}} {\bf F}_k $, the BS precoder design (\ref{General_Opt_Model}) can be transformed into
\begin{align}
	\min_{\substack{ {\bf F} } } \ 
	& \sum\nolimits_{k = 1}^K \tr \Big\lbrace 
	{\bf A}_k^\hem ( {\bf G}_k {\bf H}_{{\rm c},k} \widehat{\bf F}_k \! - \! {\bf I} ) ( {\bf G}_k {\bf H}_{{\rm c},k} \widehat{\bf F}_k \! - \! {\bf I} )^\hem {\bf A}_k \! + \! 
	{\bf A}_k^\hem {\bf G}_k {\bf R}_{{\bf n}_{\rm c}} {\bf G}_k^\hem {\bf A}_k 
	\notag \\ & \! + \! 
	\sum_{\ell \neq k} {\bf A}_k^\hem {\bf G}_k {\bf H}_{{\rm c},k} {\bf F}_{\ell} {\bf F}_{\ell}^\hem {\bf H}_{{\rm c},k}^\hem {\bf G}_k^\hem {\bf A}_k 
	\big\rbrace \notag \\
	{\rm s.t.} \ & 
	\frac{\bar{\gamma}_{{\rm D},\ell}}{P_{\rm max}} \tr \big( \widehat{\bf F}^\hem \widehat{\bf F} \big) 
	\! - \! 2 \Re \big\lbrace \tr \big( \widehat{\bf F}_{(t)}^\hem {\bf B}_{\ell} \widehat{\bf F} \big) \big\rbrace
	\! + \! \tr \big( \widehat{\bf F}_{(t)}^\hem {\bf B}_{\ell} \widehat{\bf F}_{(t)} \big) \! \le \! 0, 
	\ \ell = 1, \ldots, L,
	\label{Opt_MUMT_1}
\end{align}
where $ \widehat{\bf F}_{(t)} $ is the effective precoder at the $ t $th iteration.
It is worth noting that the optimization problem (\ref{Opt_MUMT_1}) is a standard QCQP problem and thus the optimal effective precoder $ \widehat{\bf F} $ can be found using the famous CVX toolbox \cite{cvxtool}.

After obtaining the optimized BS precoder, the RIS phase shift optimization is formulated as
\begin{align}
	\min_{ \boldsymbol{\Theta}, \boldsymbol{\theta} } \ & \sum\nolimits_{k = 1}^K \tr \big\lbrace {\bf A}_k^\hem \boldsymbol{\Phi}_{{\rm MSE},k} {\bf A}_k \big\rbrace \notag \\
	{\rm s.t.} \ \, & 
	\tr( \boldsymbol{\theta}^\hem {\bf X}_{\ell} \boldsymbol{\theta} \!\cdot\! \boldsymbol{\theta}^\hem {\bf Y}_{\ell} \boldsymbol{\theta} ) \ge \bar{\gamma}_{{\rm D},\ell}, \ \ell = 1, \ldots, L, \notag \\
	& | \theta_i | = 1, \ \ i = 1, 2, \ldots, N_{\rm R},
	\label{RIS_Opt_MUMT_1}
\end{align}
where the first inequality denotes the radar constraint. The matrices $ {\bf X}_{\ell} $ and $ {\bf Y}_{\ell} $ are defined as 
\begin{align}
	& {\bf X}_{\ell} \triangleq \big| \rho_{{\rm tg},\ell} \big| \Diag( {\bf w}_{\ell}^\T {\bf H}_{\rm BR}^* ) {\bf a}_{{\rm RIS,r},\ell}^* {\bf a}_{{\rm RIS,r},\ell}^\T \Diag( {\bf H}_{\rm BR}^\T {\bf w}_{\ell}^* ), \notag \\
	& {\bf Y}_{\ell} \triangleq \big| \rho_{{\rm tg},\ell} \big| \Diag( {\bf a}_{{\rm RIS,t},\ell}^\T ) {\bf H}_{\rm BR}^\T {\bf F}^* {\bf F}^\T {\bf H}_{\rm BR}^* \Diag( {\bf a}_{{\rm RIS,t},\ell}^* ),
	\label{Def_RIS_X_Y_MU}
\end{align}
It is shown that the RIS phase shift optimization in (\ref{RIS_Opt_MUMT_1}) has multiple non-convex quartic constraints on the RIS phase shifts $ \boldsymbol{\theta} $. Fortunately, these non-convex radar constraints have similar forms as (\ref{Radar_RIS_Quartic}), each of which can also be relaxed to the convex quadratic form as in Section~\ref{sec:single_ISAC:RIS}. On the other hand, it can be verified that the objective function in (\ref{RIS_Opt_MUMT_1}) is a quadratic function w.r.t. the RIS phase shifts $ \boldsymbol{\Theta} $. The vectorization transformation in (\ref{RIS_ADMM_theta_2}) is also applicable to problem (\ref{RIS_Opt_MUMT_1}) and the RIS phase shift design in the MUMT scenario is a quadratic programming with multiple quadratic constraints and unit-modulus constraints. Therefore, the RIS phase shifts in the MUMT scenario can also be optimized using either the proposed ADMM-based method in Section~\ref{sec:single_ISAC:RIS} or the SDR-based method.

\vspace{-4mm}

\section{Numerical Simulations}
\label{sec:simulation}

\vspace{-1.5mm}

In this section, numerical evaluations are conducted to verify both the data transmission and the target detection performance for the proposed ISAC system design.
We consider an ISAC scenario in which the BS transmits $ N = 4 $ data streams to the single user in the communication cell and detects one potential target simultaneously. Unless otherwise specified, the numbers of antennas for the BS and the user are set to be $ N_{\rm B} = 6 $ and $ N_{\rm U} = 4 $, respectively. A RIS antenna array with $ 6 \times 6 $ elements is deployed near the BS to enhance the communication and detection performance.
For the convenience of illustration, a three-dimensional Cartesian coordinate system is introduced. It is assumed that the BS, the RIS, the user, and the target are located at the coordinates of $ (0, 0, 10) $, $ (10, 50, 10) $, $ (200, -60, 0) $, and $ (-5, 35, 10) $. Concerning the sparse scattering characteristic, the Rician fading model is adopted for the communication channels, i.e., $ {\bf H} = \sqrt{\beta / (\kappa + 1)} \big( \sqrt{\kappa} {\bf H}_{\rm LoS} + {\bf H}_{\rm NLoS} \big) $, where $ \kappa $ is the Rician factor and $ \beta = \beta_0 d^{-\alpha} $ is the pathloss coefficient with channel distance $ d $ and pathloss exponent $ \alpha $. The Rician factors for the different channels are given by $ \kappa_{\rm BR} = 9 $, $ \kappa_{\rm BU} = \kappa_{\rm RU} = 0 $. The reference pathloss parameter for the unit distance $ \beta_0 $ is set to be $ \beta_0 = -30 {\rm dB} $ and the pathloss exponents are assumed to be $ \alpha_{\rm BR} = 1.8 $, $ \alpha_{\rm BU} = 3.9 $, and $ \alpha_{\rm RU} = 2.0 $, respectively. The carrier frequency for the transmit signal is $ f_{\rm c} = 10{\rm GHz} $ and transmit power of the BS is set to be $ P_{\rm max} = 20 {\rm dBm} $. Without loss of generality, the target is assumed to have a unit scattering surface area and the corresponding reflective radiation density is $ g_{\rm T}(x,y) = 1 $. During the sensing stage, the sensing time slot is given by $ T_0 = 0.05{\rm s} $ and the sampling frequency for the radar receiver is $ f_{\rm s} = 1 {\rm MHz} $. The probability of false alarm is set as $ P_{FA} = 2\% $ and the probability of detection threshold is set as $ \gamma_{\rm D} = 98\% $. The noise power at the BS side is assumed to be $ -110{\rm dBm} $. The simulations presented below are averaged over 100 independent channel realizations.

In the following simulations, the proposed ISAC system design is compared with several benchmark algorithms to examine its performance. For notational simplicity, the performance of the proposed ADMM-based ISAC system design is labeled as ``Proposed Design'', the performance of the dedicated RIS-aided communication system design is abbreviated as ``Dedicated Commun.'', and the proposed computational efficient RIS design in Section~\ref{sec:low_complex_alg} is abbreviated as ``RGD Design''. The ``SDR Design'' denotes the conventional SDR-based RIS design for the ISAC system \cite{YuanXJ2022Passive}. The proposed ISAC system design is also compared with the ISAC systems with the fixed RIS, where the ``Target Direction'' represents the ISAC system with RIS phase shifts aligned with the steering vector of the target, and the ``User Direction'' represents the ISAC system with RIS phase shifts aligned with the steering vector of the user.

\begin{figure}[!t]
	\vspace{-8mm}
	\centering
	\includegraphics[width=0.5\textwidth, trim={6mm 0mm 6mm 0mm}]{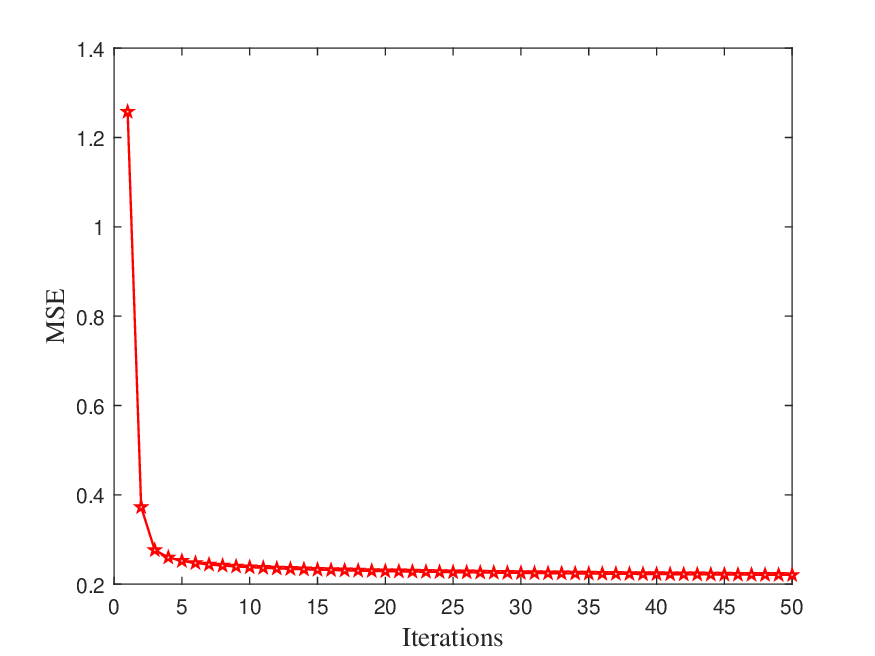}
	\vspace{-5mm}
	\caption{The convergence behavior of the MSE objective for the proposed ISAC with the user noise power of $ -110 {\rm dBm} $.}
	\label{fig_sim_1}
	\vspace{-6mm}
\end{figure}
\begin{figure}[!t]
	\centering
	\includegraphics[width=0.5\textwidth, trim={6mm 0mm 6mm 0mm}]{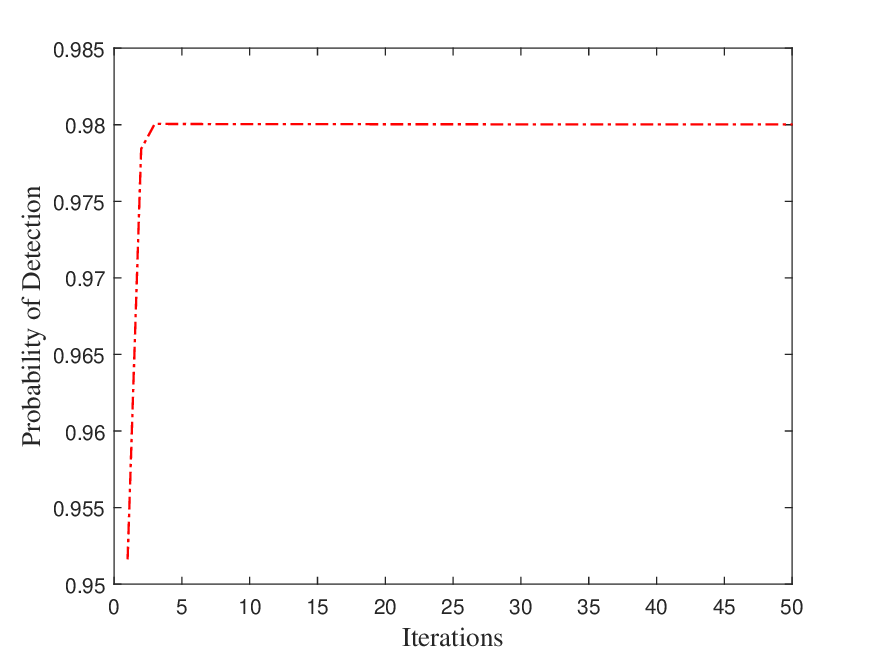}
	\vspace{-5mm}
	\caption{The convergence behavior of the radar probability of detection constraint for the proposed ISAC design with the user noise power of $ -110 {\rm dBm} $.}
	\label{fig_sim_2}
	\vspace{-8mm}
\end{figure}

\begin{figure}[!t]
	\vspace{-8mm}
	\centering
	\includegraphics[width=0.5\textwidth, trim={6mm 0mm 6mm 0mm}]{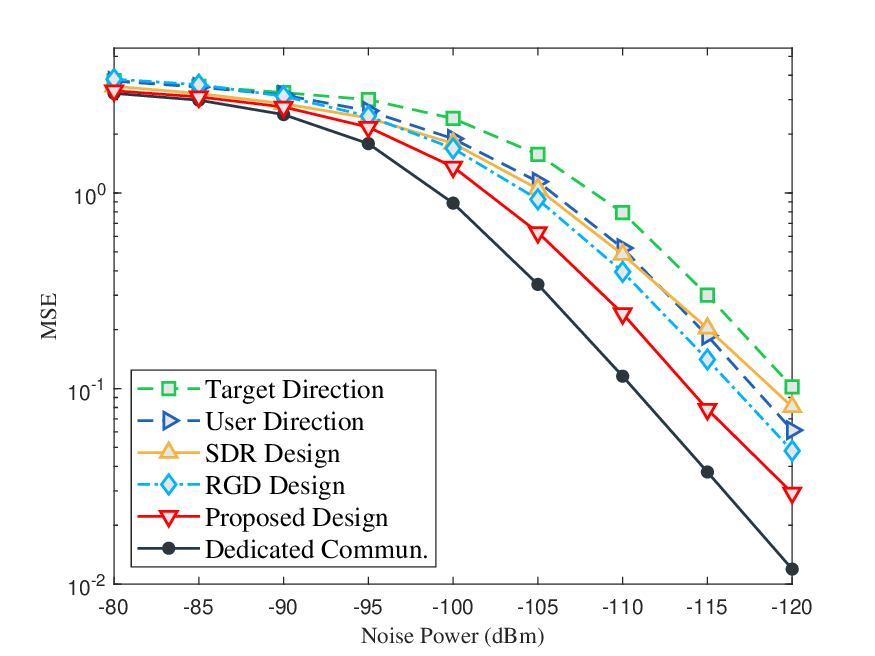}
	\vspace{-5mm}
	\caption{The MSE performance comparison of the proposed ADMM-based ISAC design with other five benchmarks under the probability of detection condition of $ \gamma_{\rm D} = 98\% $.}
	\label{fig_sim_3}
	\vspace{-6mm}
\end{figure}
\begin{figure}[!t]
	\centering
	\includegraphics[width=0.5\textwidth, trim={6mm 0mm 6mm 0mm}]{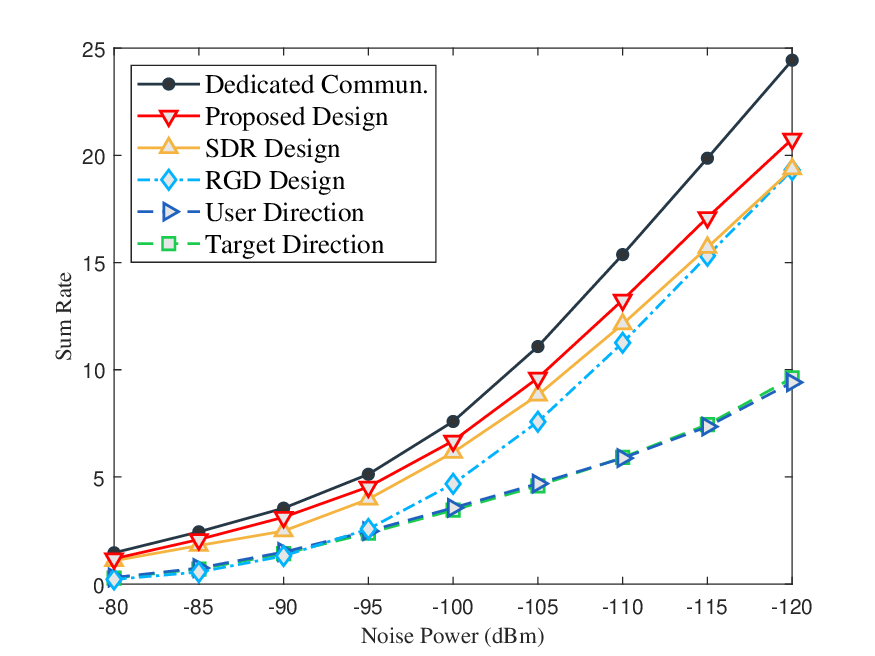}
	\vspace{-4mm}
	\caption{The sum rate performance comparison of the proposed ADMM-based ISAC design with other five benchmarks under the probability of detection condition of $ \gamma_{\rm D} = 98\% $.}
	\label{fig_sim_4}
	\vspace{-8mm}
\end{figure}

To begin with, in Fig.~\ref{fig_sim_1} and Fig.~\ref{fig_sim_2}, we aim to verify the convergence behaviors of the proposed ISAC system design from the perspectives of the objective function and constraints, respectively. In the simulation, the MSE performance is adopted with an identity weighting matrix, and the noise power level at the user side is assumed to be $ -110{\rm dBm} $. From Fig.~\ref{fig_sim_1}, it is shown that the MSE of the estimated signal monotonically decreases as the number of algorithm iterations grows and the MSE objective converges to the stationary point within 20 iterations. In Fig.~\ref{fig_sim_2}, the convergence performance for the radar detection probability constraint is testified. We clearly see from Fig.~\ref{fig_sim_2} that the radar constraint quickly converges to the required probability level after a few iterations. Thus, the radar performance can be perfectly guaranteed.

The communication performance of the proposed ISAC design is examined subsequently. Fig.~\ref{fig_sim_3} demonstrates the MSE performance comparison for the proposed design and the other five benchmarks under different signal-to-noise ratio (SNR) conditions. Based on the simulation results, it can be concluded that the proposed ISAC design achieves the best communication performance compared with other ISAC design benchmarks. However, the MSE of the proposed ISAC design is inferior to that of the traditional dedicated RIS-aided communication, since a part of the BS resource is used to detect the target rather than transmit user data. Both the Target Direction, User Direction, and the SDR-based design obtain higher transmission MSE in this scenario. On the other hand, the proposed Riemannian gradient descent design scarifies the communication performance in exchange for a lower computational complexity.

\begin{figure}[!t]
	\vspace{-8mm}
	\centering
	\includegraphics[width=0.5\textwidth, trim={6mm 0mm 6mm 0mm}]{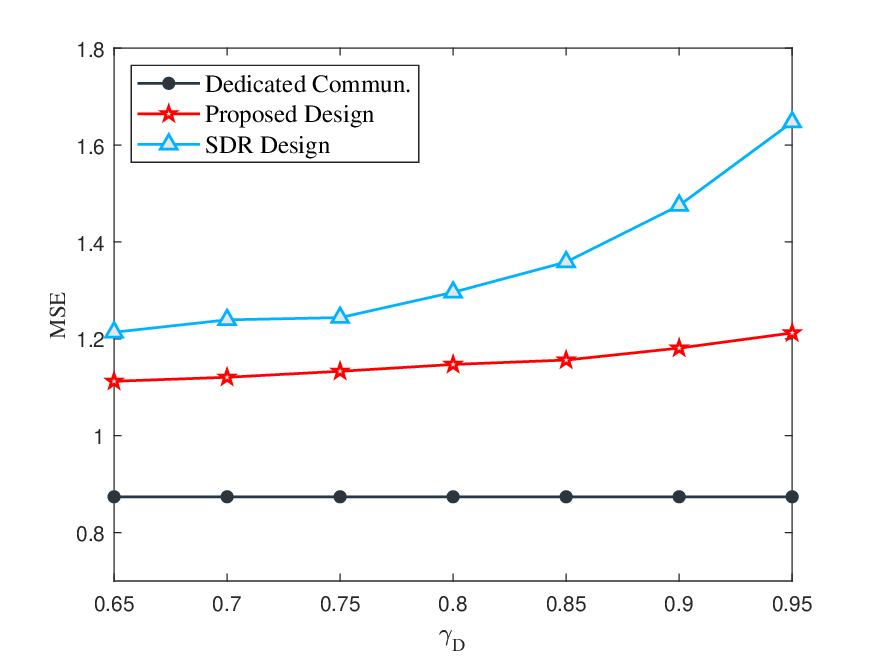}
	\vspace{-4mm}
	\caption{The normalized angle response performance of the BS for multiple ISAC designs under the probability of detection condition of $ \gamma_{\rm D} = 98\% $.}
	\label{fig_sim_7}
	\vspace{-6mm}
\end{figure}
\begin{figure}[!t]
	\centering
	\includegraphics[width=0.5\textwidth, trim={6mm 0mm 6mm 0mm}]{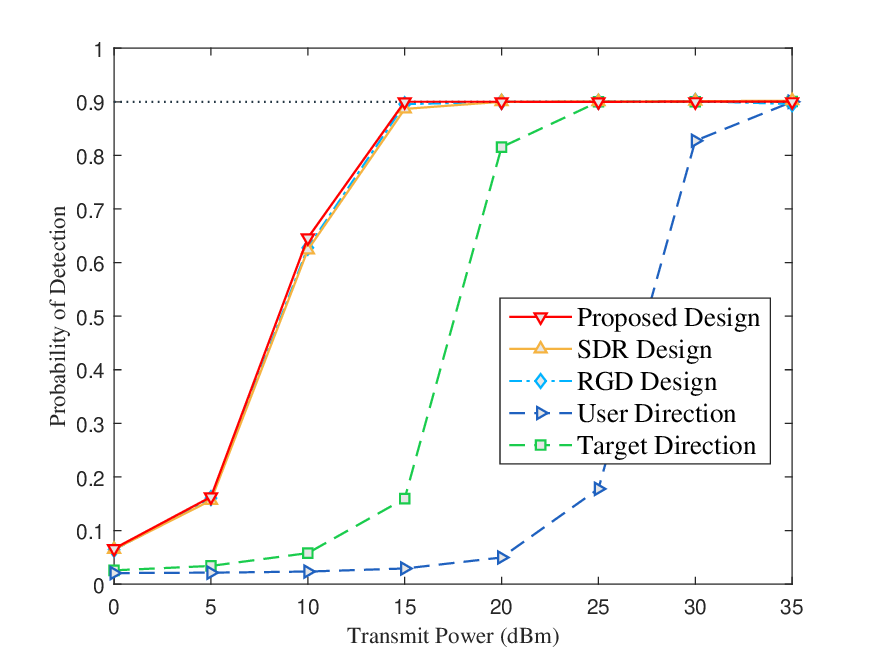}
	\vspace{-4mm}
	\caption{The detection probability comparison of the proposed ADMM-based ISAC design with other four benchmark schemes. The goal of detection probability is set as $ \gamma_{\rm D} = 90\% $ and the noise power at user node is set as $ -110{\rm dBm} $.}
	\label{fig_sim_5}
	\vspace{-8mm}
\end{figure}

Afterward, the sum rate performance is chosen as the ISAC system design criterion to testify the generality of the proposed ISAC design algorithm. In Fig.~\ref{fig_sim_4}, with the transmit power fixed, it is shown that the sum rate of the ISAC system monotonically increases as the noise power of the user decreases. The proposed ISAC design also achieves the best communication performance compared to the other ISAC designs but is still worse than the dedicated RIS-aided communication design. However, the Riemannian gradient design significantly outperforms the user direction and target direction based RIS designs. This may be attributed to the non-optimal radar beamforming which wastes too much BS resources.

Moreover, the trade-offs between the radar detection and the data transmission performance are compared under various detection probability threshold conditions in Fig.~\ref{fig_sim_7}. In the simulation, the MSE metric is adopted to illustrate the data transmission performance, and the noise power levels at the user side and the BS side are set to be $ -100 {\rm dBm} $ and $ -110 {\rm dBm} $, respectively. From Fig.~\ref{fig_sim_7}, it is demonstrated that the MSE of the communication subsystem increases with the growth of the detection probability threshold for both the proposed ADMM-based design and the conventional SDR design \cite{YuanXJ2022Passive}. This indicates the Pareto improvement phenomenon that when the ISAC system reaches the Pareto optimum, the radar subsystem can only obtain a better detection performance by sacrificing a part of the performance of the communication subsystem. It is also shown that the proposed ADMM-based design suffers less performance loss compared with the conventional SDR-based design when the ISAC system is required to improve its sensing performance. The traditional dedicated communication does not experience this trade-off, which actually forms the lower bound of the MSE performance.

\begin{figure}[!t]
	\vspace{-8mm}
	\centering
	\includegraphics[width=0.5\textwidth, trim={6mm 0mm 6mm 0mm}]{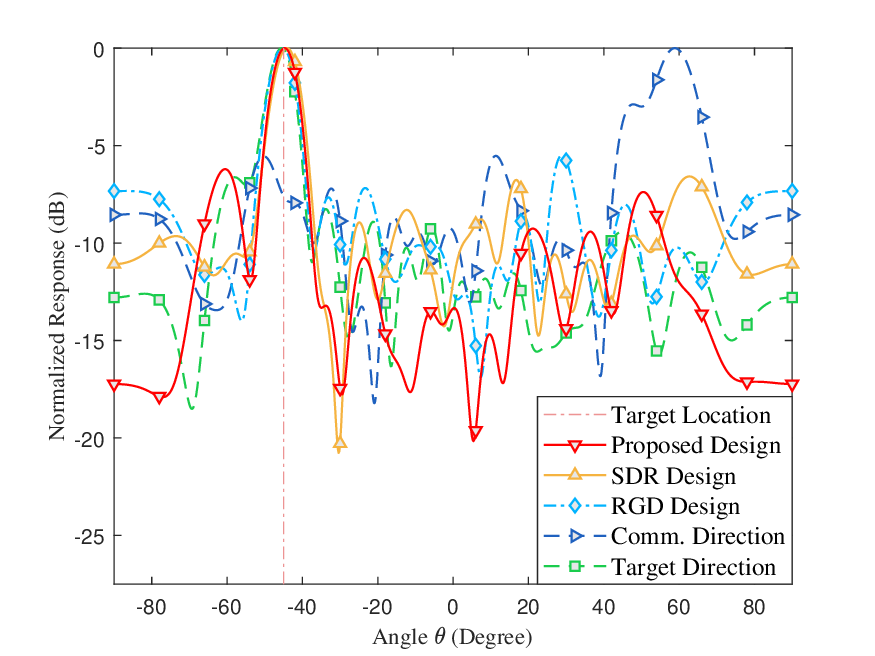}
	\vspace{-4mm}
	\caption{The normalized angle response performance of the BS for multiple ISAC designs under the probability of detection condition of $ \gamma_{\rm D} = 98\% $.}
	\label{fig_sim_6}
	\vspace{-8mm}
\end{figure}

It is worth noting that the previous two simulations illustrate the communication performance of the proposed ISAC design and the other benchmarks. As another important component of the ISAC system, the performance of the radar subsystem is non-negligible. In Fig.~\ref{fig_sim_5}, we investigate the detection probability for the radar subsystem under different transmit power. In the simulation, the noise power of the user is defined as $ -110{\rm dBm} $, and the required detection probability is set as $ \gamma_{\rm D} = 90\% $ for ease of demonstration. From Fig.~\ref{fig_sim_5}, it is demonstrated that the radar detection performance improves with the growth of the BS transmit power, since the BS is able to put more power into the radar detection process as the transmit power increases. The figure shows that the proposed ISAC design obtains better detection performance as compared to the SDR-based design and the Riemannian gradient descent based design. It is worth noting that the target direction design performs far better than the user direction design, because the beam of RIS in the user direction design does not point at the target direction. However, these two designs cannot achieve the required detection probability below the power level of $ 25{\rm dBm} $.

Finally, Fig.~\ref{fig_sim_6} compares the normalized angular response for the proposed design and other benchmarks under the BS noise power level of $ -120{\rm dBm} $. The angular response is defined as
\begin{align}
	\mathcal{P}( \vartheta, \phi ) = 
	\Vert {\bf a}_{\rm RIS,t}^\hem( \vartheta, \phi ) \boldsymbol{\Theta} {\bf H}_{\rm BR}^\hem {\bf F} \Vert_2,
\end{align}
where $ {\bf a}_{\rm RIS,t}( \vartheta, \phi ) $ and $ {\bf a}_{\rm RIS,r}( \vartheta, \phi ) $ are the steering vectors. Specifically, for the purpose of a better spatial resolution, a $ 18 \times 2 $ RIS antenna array is used in this simulation. The detection probability threshold is set as $ \gamma_{\rm D} = 98\% $ by default, and the Rician factor is given by $ \kappa = 0.23 $. According to the system setting, the angle of departure (AoD) of the RIS-to-target link is calculated as $ 45^{\circ} $. From Fig.~\ref{fig_sim_6}, it can be seen that the proposed ISAC design, the SDR based design, the Riemannian gradient based design, and the target direction design have significant responses near the target direction, but the beam direction of the proposed ISAC design is more accurately, and the proposed ISAC design also the better sidelobe leakage control. The figure also shows that the user direction design does not form a valid sensing beam to the target, which explains its worst sensing performance of in Fig.~\ref{fig_sim_5}. Moreover, it can also be noticed that the power of the sidelobes for the proposed design and other benchmarks is higher than that of the traditional MIMO radar design, which indicates that the information transmission relies on these sidelobes.

\vspace{-2mm}

\section{Conclusion}
\label{sec:conclusion}
\vspace{-1mm}

This paper investigated the general transceiver design framework for a RIS-aided MIMO-ISAC system, which is applicable to various communication performance metrics. 
Based on the problem formulation, the monotonicity of radar probability as well as the maximum power allocation criterion of the BS was proved, and the optimal BS precoder has been derived in semi-closed form. 
An iteratively ADMM-based design and a computational efficient Riemannian gradient descent design were proposed to address the quartic constrained RIS phase shift optimization.
Then, the proposed transceiver design framework has been extended to the MUMT scenario.
Numerical simulation results demonstrated the good convergence performance and superior 
communication-sensing performance of the proposed ISAC design.
However, it is also worth pointing out that our proposed design heavily relies on the knowledge of the perfect CSI for communication systems and the prior information for targets. How to obtain the information and how to develop a robust ISAC system design without the prior knowledge is our future primary interest.

\vspace{-3mm}

\appendices

\vspace{-6mm}

\section{Proof of Proposition~1}
\label{sec:appendix-A}

To show the existence of the inverse function $ P^{-1}(x) $, we first prove the monotonically increasing property of the function $ P(x) $. According to the definition of $ Q $-function, the derivative of $ P(x) $ is derived as
\begin{align}
	\frac{dP(x)}{dx} 
	& = \frac{1}{2 \sqrt{2\pi x}} e^{ - \frac{x + a^2}{2} } \Big( e^{ a \sqrt{x} } - e^{ -a \sqrt{x} } \Big).
\end{align}
Note that the domain of $ P(x) $ is real positive, i.e., $ D = \{x | x > 0\} $ and the scalar $ a $ is also positive. Based on the property of the exponential function, it can be readily concluded that $ \frac{dP(x)}{dx} > 0 $, which indicates the strictly monotonically increasing property of $ P(x) $. Based on the inversion function theorem, the inverse function of $ P(x) $ exists, and its derivative can be given by $ \frac{dP^{-1}(y)}{dy} = \big( \frac{dP(x)}{dx} \big)^{-1} > 0 $, which proves Proposition~1.
$ \hfill\blacksquare $

\vspace{-6mm}

\section{Proof of Proposition~2}
\label{sec:appendix-B}

Based on the optimal equalizer $ {\bf G}_{\rm opt} $, the precoder design problem is formulated as 
\begin{align}
	\min_{ {\bf F} } \ & f_{\rm obj} \big( ( {\bf I} + {\bf F}^\hem {\bf H}^\hem {\bf R}_{{\bf n}_{\rm c}}^{-1} {\bf H} {\bf F} )^{-1} \big) \notag \\
	{\rm s.t.} \ \ & \tr( {\bf F} {\bf F}^\hem ) \le P_{\rm max}, \notag \\
	& \tr \big( {\bf F}^\hem {\bf B} {\bf F} \big) \ge \bar{\gamma}_{\rm D}.
	\label{Opt_AP_1}
\end{align}
Then, we prove the equality of the power constraint by contradiction. Assume there exists an optimal precoder $ {\bf F}^{\star} $ in domain $ \mathcal{D} $, satisfying $ \tr ( {\bf F}^{\star} {\bf F}^{\star\hem} ) = P^{\star} \! < \! P_{\rm max} $, $ \tr ( {\bf F}^{\star\hem} {\bf B} {\bf F}^{\star} ) \! \ge \! \bar{\gamma}_{\rm D} $ and $ f_{\rm obj} \big( \boldsymbol{\Phi}_{\rm MSE}^{\star} \big) \! \le \! f_{\rm obj} \big( \boldsymbol{\Phi}_{\rm MSE} \big), \forall {\bf F} \in \mathcal{D} $, where $ \boldsymbol{\Phi}_{\rm MSE}^{\star} = ( {\bf I} + {\bf F}^{\star\hem} {\bf H}^\hem {\bf R}_{{\bf n}_{\rm c}}^{-1} {\bf H} {\bf F}^{\star} )^{-1} $ and $ \boldsymbol{\Phi}_{\rm MSE} = ( {\bf I} + {\bf F}^\hem {\bf H}^\hem {\bf R}_{{\bf n}_{\rm c}}^{-1} {\bf H} {\bf F} )^{-1} $.
However, it is easy to show that the precoder $ \bar{\bf F} = \sqrt{ P_{\rm max} / P^{\star} } \, {\bf F}^{\star} $ also lies in the feasible set, since $ \tr( \bar{\bf F} \bar{\bf F}^\hem ) \! = \! P_{\rm max} $ and $ \tr ( \bar{\bf F} {\bf B} \bar{\bf F} ) \! = \! \frac{P_{\rm max}}{P^{\star}} \tr( {\bf F}^{\star\hem} {\bf B} {\bf F}^{\star} ) > \bar{\gamma}_{\rm D} $. However, recalling the monotonicity of the performance metric in Section~\ref{sec:sys_model:comm_perform}, it is found that
\begin{align}
	f_{\rm obj} \big( \boldsymbol{\Phi}_{\rm MSE} \big) & = f_{\rm obj} \bigg( \frac{P^{\star}}{P_{\rm max}} \boldsymbol{\Phi}_{\rm MSE}^{\star} \bigg)
	< f_{\rm obj} \big( \boldsymbol{\Phi}_{\rm MSE}^{\star} \big),
\end{align}
which contradicts with the optimality assumption of precoder $ {\bf F}^{\star} $ and the proof of Proposition~2 is thus completed.
$ \hfill\blacksquare $

\vspace{-6mm}

\section{Proof of Proposition~3}
\label{sec:appendix-D}

From the optimization problem (\ref{Opt_Eq_WMSE}), it can be seen that with other variables fixed, the optimizations w.r.t. the variable $ {\bf Y} $ and $ \bar{\bf W} $ are unconstrained convex optimization problems, respectively. Thus, their optimal values can be found according to the first-order optimality condition. 
We consider the optimization of auxiliary matrix $ {\bf Y} $ at first, which is given by
\begin{align}
	\min_{ {\bf Y} } \ \, 
	& \tr \Big\lbrace \bar{\bf W} \big[ \big( {\bf Y} {\bf X} {\bf C} - {\bf I} \big) \boldsymbol{\Phi}^{-1} \big( {\bf Y} {\bf X} {\bf C} - {\bf I} \big)^\hem + {\bf Y} \boldsymbol{\Pi}^{-1} {\bf Y}^\hem \big] \Big\rbrace.
	\label{Opt_Y}
\end{align}
Let the first-order derivative of the objective function in (\ref{Opt_Y}) be zero. The optimal solution of auxiliary matrix $ {\bf Y} $ is given as
\begin{align}
	{\bf Y}_{\rm opt} = \boldsymbol{\Phi}^{-1} {\bf C}^\hem {\bf X}^\hem \big( \boldsymbol{\Pi}^{-1} + {\bf X} {\bf C} \boldsymbol{\Phi}^{-1} {\bf C}^\hem {\bf X}^\hem \big)^{-1}.
\end{align}
Based on the optimal value of $ {\bf Y} $, the optimization of weighting matrix $ \bar{\bf W} $ is written as
\begin{align}
	\min_{ \bar{\bf W} } \ \, 
	& \tr \Big\lbrace \bar{\bf W} \big[ \big( \boldsymbol{\Phi} + {\bf C}^\hem {\bf X}^\hem \boldsymbol{\Pi} {\bf X} {\bf C} \big)^{-1} \big] \Big\rbrace 
	- \log \det \big( \bar{\bf W} \big).
	\label{Opt_W}
\end{align}
As the optimization problem (\ref{Opt_W}) is still convex, the optimal solution of $ {\bf W} $ is also derived
\begin{align}
	\bar{\bf W}_{\rm opt} = \boldsymbol{\Phi} + {\bf C}^\hem {\bf X}^\hem \boldsymbol{\Pi} {\bf X} {\bf C}.
\end{align}
Substituting the optimal solutions of the auxiliary matrix $ {\bf Y} $ and the weighting matrix $ {\bf W} $ into the optimization problem (\ref{Opt_Eq_WMSE}), it can be seen that the simplified optimization of $ {\bf X} $ is the same as the original problem (\ref{Opt_WMI}), which completes the proof of Proposition~3.
$ \hfill\blacksquare $

\vspace{-6mm}

\bibliographystyle{IEEEtran}
\bibliography{IEEEabrv,RIS_ISAC_Opt}

\end{document}